\documentclass[12pt, letterpaper]{article}

\usepackage{amsmath, amsfonts, amscd, amssymb}
\usepackage{amsthm}
\usepackage[all,cmtip]{xy}

\newcommand{\aconn}{\mathcal{A}}

\newcommand{\aut}{{\mathcal{A}}ut}

\newcommand{\conn}{\mathcal{D}}

\newcommand{\finposet}{\mathcal{P}}

\newcommand{\cons}{\mathbf{C}}

\newcommand{\curv}{R}

\newcommand{\eh}{\mathcal{E}\mathcal{H}}

\newcommand{\gauge}{\mathcal{U}}

\newcommand{\gl}{\mathcal{G}\mathcal{L}}

\newcommand{\Hom}{{\mathcal{H}}om}
\newcommand{\kd}{\text{\texttt{d}}}

\newcommand{\inv}{\overleftarrow{\mathcal{P}}}

\newcommand{\modl}{\mathbf{\mathcal{E}}}

\newcommand{\omg}{\Omega}
\newcommand{\Omg}{\mathbf{\Omega}}

\newcommand{\cont}{\mathcal{C}^{0}}
\newcommand{\smooth}{\mathcal{C}^{\infty}}
\newcommand{\sconn}{\textsf{A}}

\newcommand{\struc}{\mathbf{A}}

\newcommand{\triad}{{\mathfrak{T}}}

\newcommand{\wee}{\,{\scriptstyle\wedge}\,}

\newcommand{\ym}{\mathcal{Y}\mathcal{M}}
\newcommand{\com}{\mathbb{C}}
\newcommand{\mapto}{\longrightarrow}

\newcommand{\Z}{\mathbb{Z}}
\newcommand{\R}{\mathbb{R}}

\newcommand{\unc}{\mathcal{U}}

\title{\bf Functoriality in Finitary Vacuum Einstein Gravity and Free Yang-Mills \\ Theories from an Abstract Differential Geometric Perspective\thanks{This paper is wholeheartedly dedicated to, and in loving memory of, my dearest teacher, mentor, friend, and companion in {\bf The Quest}, {\em Professor Anastasios (Tasos) Mallios}. This is an invited paper contribution to a Special Contributory Volume/Issue titled {\em Physical Geometry: Unravelling the Weave of Quantum Geometry} in memory of Professor Anastasios Mallios, edited by Dr Elias Zafiris. The paper to be submitted will also be posted at the {\bf General Relativity and Quantum Cosmology} website {\it www.arXiv.org/gr-qc} before Easter 2024. In turn, a longer version of the paper will constitute a chapter in a research monograph type of book that we have been working on, in collaboration with the late Professor Anastasios Mallios, since 2003 \cite{malrap4}.}}

\author{Ioannis Raptis\thanks{Supply \& Substitute Secondary School Teacher of Mathematics, Physics and Chemistry, Reeson Education, London, United Kingdom; email: {\it irapti11@gmail.com}}}

\date{Tuesday, 13th of February 2024}

\begin{document}

\maketitle

\pagestyle{myheadings}\markboth{\centerline {\small {\sc
{Ioannis Raptis}}}}{\centerline
{\footnotesize {\sc {Functoriality in Finitary ADG-Gravity and Yang-Mills Theories}}}}

\pagenumbering{arabic}

\begin{abstract}

\noindent{\small We continue ongoing research work \cite{malrap1, malrap2, malrap3, rap1, rap2, rap5, rap7, rap11, rap13, rap14, rapzap1, rapzap2} on applying the homological algebraic conceptual and technical machinery of Abstract Differential Geometry (ADG) towards formulating a finitary, causal and quantal version of vacuum Einstein Lorentzian gravity and free Yang-Mills theories, hitherto cumulatively referred to as \emph{ADG-Gauge Theory} (ADG-GT). In particular, we unfold, express and highlight the inherently \emph{functorial} character of ADG-GT both at the `kinematical' and at the `dynamical' level of the {\it aufbau} of the theory, although at the same time we observe that the traditional kinematics-{\it versus}-dynamics distinction becomes blurry in our ADG-theoretic approach as, in line with \cite{malrap3, malrap4, rap7, rap13, rap14, malzaf1}, we maintain that \emph{there is no pre-existent geometrical/kinematical space in the quantum deep, but rather, that physical geometrical space derives from (or is an outcome of) field dynamics}. We moreover argue that the \emph{gauge theory of the third kind} and the \emph{third quantisation} schemes that ADG-GT has been seen to support \cite{malrap3, rap13, rap14}, are also \emph{functorial} in character. Furthermore, since our inherently algebraic ADG-theoretic scheme has been seen to be manifestly \emph{background geometrical} $\smooth$-\emph{smooth spacetime manifold independent} \cite{malrap1, malrap2, malrap3, rap1, rap2, rap5, rap7, rap11, rap13, rap14}, we entertain the idea, at the `dynamical' level of functoriality, that there is both a \emph{geometric morphism} and a \emph{natural transformation} type of correspondences between the relevant Einstein and Yang-Mills field functor categories with the dynamical gauge connection and curvature sheaf morphisms implementing the homological algebraic dynamics within each, so as to further corroborate previous claims \cite{rap5, rap7, rap11, rap13, rap14} that \emph{from an ADG-theoretic perspective, ADG-gravity is an already finitistic, third quantised, $\smooth$-smooth geometrical background spacetime manifoldless, auto-dynamical and `pure gauge' field theory of the third kind}. We also cast our formal canonical sheaf cohomological Third Quantisation heuristics originally formulated in \cite{rap13} in a slightly different light so as to arrive at a new ADG-theoretic notion of {\em `Unitary' Quantal ADG-Gauge Field} which, in a tetrad of functorially and dynamically entwined structures $\mathbf{U}:=(\modl ,\conn, \aut_{\struc}\modl , \mathcal{Q})$, it subsumes under a single coherent and inseparable `unitary whole' all the four most important functorial structural traits of ADG-GT: {\em `local quantum particle states' represented by local sections of a vector sheaf $\modl$, their `dual-complementary' functorial ADG-gauge field dynamics generated by an algebraic $\struc$-connection $\conn$, the latter's local gauge invariance of the 3rd kind encoded in the principal structure sheaf $\aut_{\struc}\modl$ of $\modl$'s automorphisms, and the dual particle-field canonical-type of 3rd quantisation, represented by the functorial morphism $\mathcal{Q}$ between the sheaf categories involved}. At the end, we give a subjective (:from this author's viewpoint) account of certain key ideas, concepts and seminal mathematical results in the past that significantly motivated Professor Mallios to develop ADG, and how these ideas subsequently inspired this author to apply it to a finitistic and quantal theoretical scenario for Vacuum Einstein Gravity and Free Yang-Mills theories. Throughout the second half of the paper, we recall and analyse several pertinent quotes that Professor Mallios and this author used to repeatedly discuss and scrutinise in the course of the late nineties and early noughties, during endless late night discussions over good food and wine at our favourite tavern, fittingly called {\em Algebra}, in Paleo Psychiko, Athens, Greece. Addendum 1 at the end recalls an important and telling early interaction that this author enjoyed with Professor Mallios at the end of last century. Addendum 2 at the end discusses {\em the importance of using poetic language}, plus imaginative and heuristic novel terminology, both of which emanate from the novel mathematical concepts, structures and techniques of ADG, in order to address, interpret and formulate new theoretical concepts and calculational techniques in the wildly speculative, glaringly non-intuitive and largely uncharted landscape of Quantum Gravity. An Appendix, defining, describing and explaining all the new ADG-theoretical concepts, concludes the paper.}
\vskip 0.1in



\end{abstract}

\newpage

\section{Technical Prolegomena: A Brief History of Finitary ADG-Gravity and Yang-Mills Theories with an Emphasis on the Functorial Character of our Concepts, Methods and Constructions}

In this section we give a brief account of the main milestones reached along our way towards arriving at a purely algebraic, finitistic, causal and quantal theory of spacetime, gauge theories and gravity. As we outline the main results, we highlight and emphasise the homological algebraic (:category-theoretic), and especially \emph{functorial}, nature of our basic concepts, structures and methods of their use in various constructions and associated (abstract differential geometric) calculations (:Differential Calculus), while {\em all this is accomplished purely algebraically, manifestly without any recourse to or dependence on a background $\smooth$-smooth geometrical base spacetime manifold}.

\subsection{Finitary Substitutes of Continuous Spacetime Manifolds, their Incidence Algebras, and the Finitary Sheaves Thereof: `Kinematical' Functoriality}

\subsubsection{Sorkin's Finitary Posets} 

Our journey begins with Sorkin's `prophetic' \emph{finitary substitutes of continuous spacetime manifolds} \cite{sork0}\footnote{Refer to this paper for various mathematical concepts, structures and technical definitions thereof.}. In that paper, with every \emph{locally finite (:finitary) open cover} $\gauge_{i}=(U_{i})$ of a (real) topological ($\cont$) spacetime manifold $M$, Sorkin assigns a so-called \emph{finitary partially ordered set} (finposet) $\finposet_{i}$:

\begin{equation}\label{eq1}
\gauge_{i}\longrightarrow\finposet_{i}\;\;\; 
\end{equation}

\noindent The collection $\inv =(\finposet_{i})_{i\in I}$ of such finposets is seen to constitute a so-called \emph{inverse} or \emph{projective} system, or \emph{net}, of posets, which is seen to have a \emph{projective limit space} effectively homeomorphic to the continuous $\cont$-manifold $M$.\footnote{The index $i\in I$ in $\finposet_{i}$ is the so-called \emph{refinement net index}, whereby $\gauge_{i}\prec\gauge_{j}$ (reads: the open cover $\gauge_{j}$ is finer than the open cover $\gauge_{i}$, and conversely, $\gauge_{i}$ is coarser than $\gauge_{j}$) if $\gauge_{j}$ has `smaller' and more numerous open sets than $\gauge_{i}$. The aforementioned inverse limit now reads: \emph{the $\cont$-manifold is recovered at the inverse limit of infinite refinement of the open covers} $\gauge_{i}$s \emph{in the net} (as $i\rightarrow\infty$).}

Three bullet points must be emphasised here in connection with (\ref{eq1}) from \cite{sork0}:

\begin{itemize}

\item First, Sorkin's original intuition that every geometrical point of a point-manifold is an ideal, operationally unrealistic and physically untenable plus problematic (:singular), dimensionless and structureless object, that better be `smeared' and blown-up by `enlarged' open sets (neighbourhoods) about it.\footnote{Let it be noted here that a pointed background geometrical spacetime continuum is a problematic, pathological structure that is arguably responsible both for the singularities plaguing General Relativity (GR) and for the pestilential non-renormalisable unphysical infinities marring Quantum (Gauge) Field Theories (QFT) of matter.}  The open sets and their set-theoretic algebra are the carriers of the manifold's topology and its continuity, not its ideal points.\footnote{In the ensuing discussion, by introducing \emph{sheaf theory}, we will refine this statement even more.}

\item Sorkin's original idea strongly resonates with Grothendieck's pioneering idea to categorically abstract and generalise pointed topological spaces to \emph{pointless} ones called \emph{sites} by abstracting from the usual topological open covers like in $\inv =(\gauge_{i})=(\finposet_{i})_{i\in I}$ to \emph{families} (=\emph{sieves}) \emph{of covering arrows in a category} defining a \emph{Grothendieck topology} on the category \cite{maclane1, maclane2, macmo}.\footnote{A \emph{site} is by definition a category endowed with a Grothendieck topology.}

\item Once we have done away with the pointed geometrical manifold continuum, we can work further with their poset substitutes $(\finposet_{i})$ and build on them. That is what we do next.

\end{itemize}

\noindent {\bf First encounter with `Kinematical' Functoriality.} Before we go on to work with the finitary posets, we catch a first glimpse of \emph{functoriality} of our constructions. Let $\mathcal{T}_{i}=\mathrm{span}\{ U:\; U\in\gauge_{i}\}$ be \emph{the topology `spanned' or generated by arbitrary unions and finite intersections of the open sets in each locally finite open covering} $\gauge_{i}$. Then, a continuous map

\begin{equation}\label{eq2}
f:\; \mathcal{T}_{i} \longrightarrow \mathcal{T}_{j}
\end{equation}

\noindent induces (:maps functorially to) a \emph{poset morphism} $\hat{f}:\; \finposet_{i}\rightarrow\finposet_{j}$,\footnote{By definition, \emph{a poset morphism is a partial order preserving map}.} such that the following diagram commutes:

\begin{equation}\label{eq3}
\xymatrix{
\mathcal{T}_{i} \ar[d] \ar[r]^{f} & \mathcal{T}_{j} \ar[d] \\
\finposet_{i} \ar[r]_{\hat{f}} & \finposet{j} }
\end{equation}

\subsubsection{\bf Differential Incidence (Rota) Algebras of Finitary Posets and their Simplicial Complexes} 

From \cite{zap0, zap1, rapzap1, rapzap2} we read that with every finitary poset $\finposet{i}$ one can straightforwardly associate a \emph{finitary simplicial complex} $\mathcal{S}_{i}$, the so-called \emph{\v{C}ech-Alexandrov nerve of the underlying finitary open covering}, \'a-la \v{C}ech Homology:

\begin{equation}\label{eq4}
\mathcal{S}_{i}:\: \finposet_{i}\longrightarrow\mathcal{S}_{i}
\end{equation}

\noindent {\bf Second encounter with `Kinematical' Functoriality.} The mapping $\mathcal{S}$ above is also \emph{functorial} in the sense that a finitary poset morphism $p_{ij}:\; \finposet_{i}\rightarrow\finposet_{j}$ as before, induces (:maps functorially to) a \emph{simplicial mapping}: $\hat{s}_{ij}:\; \mathcal{S}_{i}\rightarrow\mathcal{S}_{j}$, so that the following diagram commutes:

\begin{equation}\label{eq5}
\xymatrix{
\finposet_{i} \ar[d]_{\mathcal{S}_{i}} \ar[r]^{p_{ij}} & \finposet_{j} \ar[d]^{\mathcal{S}_{j}} \\
\mathcal{S}_{i} \ar[r]_{\hat{s}_{ij}} & \mathcal{S}_{j} }
\end{equation}

\noindent {\bf Third encounter with `Kinematical' Functoriality: Incidence (Rota) Algebras.} More importantly for our considerations here, we read from \cite{zap0, zap1, rapzap1, rapzap2} that with every finitary poset $\finposet_{i}$ (or equivalently, with every finitary \v{C}ech simplicial complex $\mathcal{S}_{i}$) one can naturally associate a so-called \emph{incidence (Rota) algebra} $\Omega_{i}$ over the complex numbers $\com$, as follows: 

\begin{equation}\label{eq6}
\mathcal{R}_{i}:\;\finposet_{i}\rightarrow\Omega_{i}
\end{equation}

\noindent Like in the case of the simplicial mapping of finitary posets $\mathcal{S}$, the mapping $\mathcal{R}$ induces an \emph{incidence algebra homomorphism} $\hat{r}$ in the following functorial commutative diagram sense:\footnote{{\it Mutatis mutandis} for the finitary  \v{C}ech simplicial complexes and the incidence Rota algebras thereof: their correspondence is manifestly functorial \cite{zap0, zap1, rapzap1, rapzap2}.}

\begin{equation}\label{eq7}
\xymatrix{
\finposet_{i} \ar[d]_{\mathcal{R}_{i}} \ar[r]^{p_{ij}} & \finposet_{j} \ar[d]^{\mathcal{R}_{j}}\\
\Omega_{i} \ar[r]_{\hat{r}_{ij}} & \Omega_{j} }
\end{equation}

\subsubsection{\bf Fourth encounter with `Kinematical' Functoriality: Gel'fand Duality} 

Without going into detailed technicalities here, we read from \cite{zap0, zap1, rapzap1, rapzap2} that one can go the other way around and extract from the finitary incidence algebras a topological space, endowed with a so-called \emph{Rota topology}, by considering \emph{irreducible representations of the incidence algebras}, the \emph{kernels} of which correspond to \emph{primitive ideals} in the algebras. In turn, the set of primitive ideals, the so-called \emph{spectrum of the algebra} $\mathrm{Spec}(\Omega)$, now regarded as a \emph{generalised, `blown up point set'}, is readily endowed with a Rota topology in such a way that \emph{the incidence algebra homomorphisms in} ${\hat{r}_{ij}}$ \emph{lift to continuous maps in the respective Rota topologies}.

This is another instance of \emph{the functoriality of our constructions} and it corresponds to a finitary version of \emph{Gel'fand Duality} according to which, very broadly speaking, from an algebraic structure $A$ one can extract a `geometrical space' $\mathrm{Spec}(A)$ carrying a `natural continuity' (:a functorially imposed topology on it).\footnote{Furthermore, and again very broadly speaking, there is the \emph{Gel'fand Representation theorem} that ensures that the $\com$-algebra of continuous complex valued functions on $\mathrm{Spec}(A)$ is naturally equivalent to the (complex) algebra $A$ that one started with. This remark will prove crucial in the sequel when we recount the introduction of \emph{finitary spacetime sheaves}.}

\subsubsection{\bf Finitary Incidence (Rota) Algebras as Finitary Differential Algebras/Modules} 

We read again directly from \cite{rapzap1, rapzap2, malrap2} that \emph{the finitary incidence algebras} $\Omega_{i}$ are $\Z_{+}$\emph{-graded discrete differential algebras/modules of finite rank},\footnote{See also \cite{dimu1, dimu3, dimu2} for an early study of such differential spaces.} as follows:

\begin{equation}\label{eq8}
\Omega = \bigoplus_{n\in\Z_{+}}\Omega^{n}=\Omega^0 \oplus \Omega^1 \oplus
 \ldots =\struc\oplus\mathcal{R}
\end{equation}

\noindent The $\Omega^{n}$s above are seen to be the reticular analogues of the usual linear spaces of $n$-grade (Grassmann exterior) differential forms \cite{malrap2} on a $\smooth$-smooth manifold. The grade $0$ commutative linear subalgebra $\struc=\Omega^{0}$ is the discrete analogue of the algebra of (smooth) functions ($0$-forms) on the continuum, while $\mathcal{R}=\bigoplus_{n\geq 1}\Omega^{n}$ serves as the $\struc$-module of discrete differential forms on it.

Furthermore, we witness in \cite{malrap2, malrap3} that there is a discrete analogue of the (flat) Cartan-K\"ahler (exterior) differential $d$ operator: 

\begin{equation}\label{eq9}
d:\; \Omega^{n}\longrightarrow\Omega^{n+1}
\end{equation}

\noindent that is a \emph{linear map} and it obeys the \emph{Leibniz rule}.

\subsubsection{\bf The Differential Caveat: Finitary Spacetime Sheaves of Incidence Algebras and Preliminary Vibes of ADG} 

As soon as this author realised that \emph{the incidence algebras encode not only topological, but also differential geometric, information in their structure}, the next tenable position would be to somehow make them \emph{(dynamically) variable}, thus he envisaged to employ in the longer run the full ADG-theoretic panoply towards formulating a finitary and quantal version of Gravity and Gauge Theory. 

A first step to that end would be to `\emph{sheafify}' them; that is, to consider \emph{sheaves} thereof.\footnote{Along very similar lines of thought, the reader should refer to the Introduction of \cite{malzaf1} to read how the notion of a \emph{sheaf} comes hand in hand with the notion of \emph{variable structure}.} Thus \emph{finitary spacetime sheaves} \cite{rap2} $\modl_{i}$ of incidence algebras\footnote{Now the locally finite posets being interpreted as \emph{causal and quantal} versions of \emph{Sorkin's causal sets} \cite{bomb87, sork1, sork2, sork4}, as expounded in \cite{rap1}.} over Sorkin's finitary poset discretisations were born.

\noindent  {\bf Fifth encounter with `Kinematical' Functoriality: Sheafification.} That is, this author realised in \cite{rap2} that the mapping $\mathcal{R}_{i}:\;\finposet_{i}\rightarrow\Omega_{i}$ in \ref{eq6} above is actually a \emph{contravariant functor}\footnote{Effectively, the definition of a \emph{presheaf} \cite{bredon}.} which, when subjected to suitable compatibility (glueing) conditions, it can be promoted to a \emph{local homeomorphism} (between the corresponding covering topologies $\mathcal{T}_{i}$ generated by the finitary open covers $\gauge_{i}$), the very definition of a sheaf \cite{bredon}.

All in all, when the resulting finitary spacetime sheaves $\modl_{i}$ have $\Omega_{i}$s in their stalks, they were recognised as being the reticular analogues of Mallios's \emph{vector sheaves}.\footnote{By definition, Mallios's vector sheaves are locally free (differential) $\struc$-modules of finite rank $n$ \cite{mall1, mall2, mall4}. That is, locally for every open set  $U\in\gauge{i}$ in an open covering (:set of local gauges $\gauge_{i}$) of the base topological space $X$ of a \emph{$\com$-algebraized space $(X,\struc)$}, one has by definition the following $\struc|_{U}$-isomorphisms: $\modl|_{U}=\struc^{n}|_{U}=(\struc|_{U})^{n}$ and, concomitantly,
the following equalities section-wise: $\modl(U)=\struc^{n}(U)=\struc(U)^{n}$ (with $\struc^{n}$ the $n$-fold Whitney sum of $\struc$ with itself).} Hence, the whole enterprise of applying Mallios's Abstract Differential Geometry (ADG) to to a finitistic, causal and quantal version of Lorentzian vacuum Einstein Gravity and free Yang-Mills (gauge) theories of matter commenced.

In this line of thought, in \cite{malrap1, malrap2, malrap3} we defined \emph{finitary differential triads}, as the following triplets:

\begin{equation}\label{eq10}
\mathcal{T}_{i}:=(\struc_{i},d,\Omega_{i})
\end{equation}

\noindent which are the `discrete' analogues of the ADG-theoretic differential triads $(\struc,\partial ,\modl\simeq\Omega)$.\footnote{Refer to \cite{mall1,mall2,mall4,malrap1,malrap2,malrap3} for more detailed definitions, further interpretational discussion and relevant results.}

\subsubsection{Sixth encounter with `Kinematical' Functoriality: The Category of Differential Triads} 

In \cite{pap1, pap2}, Papatriantafillou observed that the ADG-theoretic differential triads form a very homologically rich category: \emph{the category of differential triads}, whose objects are differential triads and whose arrows are differential structure preserving {\em sheaf morphisms}.

What is very interesting for us here, as observed in \cite{malrap1, malrap2, malrap3, rap7, rap14}, is that \emph{there is a contravariant functor between the category of finitary posets and the category of finitary differential triads of incidence algebras}. As a result, as the inverse system of Sorkin's finitary posets was seen to possess an inverse (or projective) limit space, the corresponding, categorically dual, inductive system of finitary incidence algebras was seen to form a direct (or inductive) limit space, consistent and in agreement with Papatriantafillou's results in \cite{pap2}.\footnote{Subsequently, this observation was crucial in our idea of promoting our category of finitary differential triads into a \emph{topos-like structure} \cite{macmo, rap7, rap14}, having (at least finite) categorical limits (direct/inductive) and colimits (inverse/projective).}

\subsection{Enter ADG: `Dynamical' Functoriality}

(\underline{\bf Note:} Henceforth in this paper, all our finitary considerations, constructions and results presented thus far carry on {\it mutatis mutandis} to the general ADG-theoretic constructions. Thus, the finitary case is obtained from the general ADG-theory \cite{mall1, mall2, mall4} simply by adjoining a finitarity index-subscript `$i$' to all the ADG-symbols and constructions.\footnote{For instance, a general open cover $\gauge$ of the background topological space $X$ in ADG (:there coined {\em open coordinate gauge of $X$}), becomes $\gauge_{i}$ in our finitary domain of the theory. Similarly, as we shall see next, the finitary version of the Cartan-K\"ahler differential operator $d$ or the general ADG $\struc$-connection operator $\conn$, become $d_{i}$ and $\conn_{i}$ respectively in our finitary realm {\em without any loss of generality whatsoever} \cite{malrap1, malrap2, malrap3, rap5, rap7, rap11, rap13, rap14}.})

With the identification of the aforementioned finitary sheaves of quantum causal sets as reticular versions of \emph{vector sheaves} in ADG, we swiftly moved on to mathematically model \emph{dynamical variations} thereof.

\subsubsection{Briefly revisiting $\struc$-connections in ADG}

To that end, we readily appreciated that the Cartan-K\"{a}hler differential operator in (\ref{eq9}) is a special example of an ADG-theoretic \emph{connection} on the sheaves of `differential forms' that it functorially acts as a \emph{sheaf morphism}, albeit a \emph{flat connection} \cite{mall1, mall2, mall4, malrap1, malrap2, malrap3}.

In order to dynamically vary the quantum causal sets that dwell as (germs of) local sections in the stalks of the aforementioned sheaves $\modl_{i}$, we need to `gauge'\footnote{That is, we need to localise and relativise differential changes relative to arbitrary sets of (covering) open gauges $(\gauge_{i})$ \cite{mall1, mall4, malrap1, malrap2, malrap3}.} the flat $d$ to a more general (:curved) connection $\conn$, 

\begin{equation}
d\longrightarrow\conn \label{eq11}
\end{equation}

\noindent  \emph{which is also defined functorially as an $\struc$-linear, Leibnizian sheaf morphism}, acting on the relevant module sheaves as follows:

\begin{equation}\label{eq12}
\conn:~\modl\mapto\modl\otimes_{\struc}\Omg\cong
\Omg\otimes_{\struc}\modl\equiv\Omg(\modl)
\end{equation}

\noindent With the introduction of $\conn$ upon localising or `gauging' the flat differential $d$ relative to a set of local open gauges $\gauge_{i}$, the latter acquires locally an additional term---the so-called \emph{gauge vector field potentials'} term $\aconn$,\footnote{Traditionally, in the Classical Differential Geometry (CDG) $\smooth$-smooth manifolds $M$ used by Physics \cite{gosch}, the term \emph{connection} is normally reserved for the so-called \emph{gauge vector field potentials} $\aconn_{\mu}^{i}$ (with $\mu$ an external spacetime index, and $i$ an internal gauge symmetry index). On the other hand, from an ADG-theoretic perspective, the denomination \emph{connection} is a `holistic', `unitary' one, pertaining to $\conn$ as a whole, and not referring to its `contingent' local split by a choice of gauge $\gauge_{i}$ as in (\ref{eq13}) above. Read on.} as follows:

\begin{equation}
d\longrightarrow\conn|_{U\in\gauge_{i}}= d+\aconn|_{U} \label{eq13}
\end{equation}

\noindent As alluded to in the last footnote, from an ADG-theoretic point of view, the connection $\conn$ is viewed as a `unitary', autonomous dynamical entity \cite{malrap3, rap13, rap14}, regardless of its local gauge split as in (\ref{eq13}). Which brings us to arguably the most important ADG-theoretic definition.

\subsubsection{\bf First encounter with `Dynamical' Functoriality: ADG-theoretic Fields} 

In ADG, a \emph{dynamical field} is defined as the following pair:

\begin{equation}
\mathcal{F}:=(\modl,\conn) \label{eq14}
\end{equation}

\noindent That is to say:

\begin{quotation}

\noindent \emph{A dynamical field is a pair consisting of a vector sheaf $\modl$, localised on an in principle arbitrary $\com$-algebraized space $(X,\struc)$, and a connection $\conn$ acting functorially on its (local) sections as a sheaf morphism.}

\end{quotation}

\noindent Four things to highlight here in connection with the fundamental definition above:

\begin{itemize}

\item An ADG-field consists of both the source and the agent of dynamical variability---the connection $\conn$, and the recipient of the agent's dynamical action---the vector sheaf $\modl$, as {\em an autonomous and indivisible/inseparable unit}.

\item From a geometric (pre)quantization perspective, the local sections of $\modl$ correspond to quantum particle states \cite{mall5, mall6, mall4, mall13, malrap2, malrap3, rap13, rap14}. If $\modl$ is a \emph{line sheaf} $\mathcal{L}$ (:a vector sheaf of rank $n=1$), its local sections represent the quantum particle states of a boson like the `photon', hence the ADG-field $\mathcal{F}_{Max}=(\mathcal{L},\conn)$ is coined the \emph{Maxwell field}. More general \emph{Yang-Mills} ADG-fields are represented as connections $\conn$ on vector sheaves $\modl$ of rank $n>1$. ADG-theoretically, we represent them by the pair: $\mathcal{F}_{YM}=(\modl ,\conn)$. Finally, the gravitational connections constitute ADG-theoretic \emph{Einstein fields}: $\mathcal{F}_{Einst}=(\modl ,\conn)$.

\item It is important to stress here that, in a very technical and rigorous sense, \emph{the vector sheaves $\modl$ correspond to the associated or representation sheaves of the principal group sheaf $\aut(\modl)$ of the reversible endomorphisms (:the automorphisms) of $\modl$} \cite{mall1, vas1, vas2, vas3, vas4, mall4}.\footnote{We recall from \cite{mall1, malrap3, mall4}
that for any vector sheaf $\modl$, ${\modl}nd\modl\equiv{\mathcal{H}}om_{\struc}(\modl ,\modl)\cong\modl\otimes_{\struc}\modl^{*}=\modl^{*}\otimes_{\struc}\modl$, so that $\aut(\modl)\simeq\modl nd(\modl)^{*}$. It follows that, for a choice of local open gauges $U\in\gauge_{i}$, $\aut(\modl)_{U}\equiv\aut(\modl)(U)\simeq\modl nd(\modl)^{*}(U)=M_{n}(\struc_{U})\equiv M_{n}(\struc(U))$, the \emph{non-commutative gauge structure group sheaf} of $(n\times n)$-matrices having for entries local sections in the structure algebra sheaf $\struc$: $\Gamma(U,\struc)=\struc_{U}\equiv\struc(U)$.} In turn, $\aut(\modl)$ \emph{is the local relativity and gauge invariance structure group sheaf of the functorial dynamics effectuated by the connection sheaf morphism $\conn$ acting dynamically on (the local sections of)} $\modl$.

\item The Maxwell $\mathcal{F}_{Max}=(\mathcal{L},\conn)$, Yang-Mills $\mathcal{F}_{YM}=(\modl ,\conn)$ and Einstein $\mathcal{F}_{Einst}=(\modl ,\conn)$ ADG-fields can be organised, as objects, into respective {\em categories} with categorical sheaf morphisms as arrows between them, coined: the {\em Maxwell Category} $\mathcal{T}_{Max}$, the {\em Yang-Mills Category} $\mathcal{T}_{YM}$ and the {\em Einstein Category} $\mathcal{T}_{Einst}$ categories \cite{mall4, mall14, mall13}. In contradistinction to the `flat', ungauged and `static-kinematical' categories of differential triads that we alluded to earlier \cite{pap1, pap2, mall4}, these three categories are `dynamical' in character, in the sense that object-fields in them obey and satisfy certain dynamical laws of motion\footnote{The Maxwell, Yang-Mills and Einstein dynamical (differential) equations, which in turn derive from Lagrangian variation of corresponding action functionals. Read on.} and the $\struc$-functoriality thereof corresponds to the $\struc$-invariance and the $\aut_{\struc}\modl$ local gauge/generalised coordinates' invariance of the respective dynamical laws of motion \cite{mall14, mall13}.\footnote{We shall return to discuss further the categorical implications and the deeper physical interpretation of the generalised ADG-theoretic conception of local gauge invariance as $\struc$-invariance and the $\otimes_{\struc}$-functoriality of the ADG-theoretic gauge dynamics in the sequel.}

\noindent We close this subsection with an important observation regarding the three $ADG$-field categories defined above:

\begin{quotation}

\noindent All three aforementioned ADG-field categories, the {\em Maxwell} $\mathcal{T}_{Max}$, the {\em Yang-Mills} $\mathcal{T}_{YM}$, and the {\em Einstein} category $\mathcal{T}_{Einst}$, are by definition \emph{functor categories} \cite{macmo}.\footnote{That is to say, \emph{the objects in those categories are sheaf morphisms, while the arrows between them are themselves functors}. It follows, that if there are functorial correspondences between them and other functor categories, these correspondences will be some kind of \emph{natural transformations}, and especially, some kind of \emph{geometric morphisms} \cite{macmo}. As we will see in the sequel, of special interest to us will be a geometric morphism asssociated with the \emph{$\otimes_{\struc}$-$\mathrm{Hom}$ adjunction}, which effectuates a kind of natural transformation between ADG-field and ADG-curvature space categories. Read on.}

\end{quotation}

\end{itemize}

\subsubsection{\bf Second encounter with `Dynamical' Functoriality: $\struc$-invariance is a generalised, functorial form of gauge invariance of the ADG-field dynamics} 

The discussion above brings us to the all-important issue of gauge covariance, local gauge invariance and their intimate relation to the basic ADG-theoretic notion of $\aconn$-invariance.\footnote{As it has also been observed in past publications \cite{malrap1, malrap2, malrap3}, in our work we use the symbol $\struc$ for the \emph{structure sheaf of algebras of generalised ADG-theoretic coordinate functions}, as opposed to $\aconn$ used throughout Mallios's work \cite{mall1, mall2, mall4, mall14, mall13}, as we have reserved the symbol $\aconn$ for local \emph{Einstein gravitational and Yang-Mills local gauge potentials} as in (\ref{eq13}).} In this subsection, we will focus only on \cite{non-abelian} Einstein and Yang-Mills ADG-fields on higher rank vector sheaves, leaving the \emph{abelian} case of Maxwell fields on line sheaves to their exhaustive treatment in the monograph references \cite{mall1, mall4}.\footnote{The epithets \emph{abelian} and \emph{non-abelian} above pertain, as in the usual theory \cite{gosch}, to the structure gauge groups being \emph{commutative} and \emph{non-commutative}, respectively. Indeed, the principal group sheaves associated with the line sheaves $\mathcal{L}$ of the ADG-theoretic Maxwell fields $\mathcal{F}_{Max}$ above carry abelian unitary (:$\equiv U_{1}(\struc)$) groups in their stalks, while as we saw couple of footnotes earlier, the structure group sheaf $\aut_{\struc}\modl$ of dynamical automorphisms of the vector sheaf $\modl$ of, say, the Einstein field  $\mathcal{F}_{Einst}$ is locally homomorphic to $M_{n}(\struc(U))$, \emph{which is manifestly noncommutative}.}

\noindent{\bf The $\struc$-Functoriality of the Curvature of an $\struc$-Connection: $\curv$ is an $\otimes_{\struc}$-tensor.} To that end, we first recall from \cite{mall1, mall2, mall4, malrap1, malrap2, malrap3} the general \emph{functorial} ADG-theoretic definition of the curvature $\curv$ of an $\struc$-connection $\conn$ as the following \emph{$\struc$)-morphism of $\struc$-modules}:\footnote{With a vector sheaf $\modl$, as explained before, regarded as a\emph{sheaf of differential $\struc$-modules, with structure sheaf $\struc$, that is locally isomorphic to $\struc^{n}(U)$}.}

\noindent We first define the {\em 1st prolongation of $\conn$} to be the following
$\cons$-linear vector sheaf morphism:

\begin{equation}\label{eq15}
\conn^{1}:~\Omg^{1}(\modl)\mapto\Omg^{2}(\modl)
\end{equation}

\noindent satisfying section-wise relative to $\conn$: 

\begin{equation}\label{eq16}
\conn^{1}(s\otimes t):=s\otimes\kd t-t\wee\conn
s,~(s\in\modl(U),t\in\Omg^{1}(U),U~\mathrm{open~in}~X)
\end{equation}

We are now in a position to define the curvature $\curv$ of an
$\struc$-connection $\conn$ by the following triangular commutative diagram:

\begin{equation}\label{eq17}
\xymatrix{
\modl \ar[d]|{\curv(\conn)=\conn^{1}\circ\conn} \ar[r]^{\conn}
& \Omg^{1}(\modl)\equiv\modl\otimes_{\struc}\Omg^{1} \ar[dl]^{\conn^{1}} \\
\Omg^{2}(\modl)\equiv\modl\otimes_{\struc} \Omg^{2} }
\end{equation}

\noindent from which we read directly that:

\begin{equation}\label{eq18}
\curv\equiv \curv(\conn):=\conn^{1}\circ\conn
\end{equation}

\noindent Therefore, any time we have the $\cons$-linear morphism
$\conn$ and its prolongation $\conn^{1}$ at our disposal, we can
define the curvature $\curv(\conn)$ of the connection
$\conn$.\footnote{In connection with (\ref{eq18}), one can justify
our earlier remark that the standard Cartan-K\"{a}hler (exterior) differential operator $d\equiv d^{0}$ is a {\em flat} type of connection, 
since: $\curv(d)=d\circ d\equiv d^{2}=0$, which is secured by the well known \emph{nilpotency} of the
usual Cartan-K\"{a}hler (exterior) differential operator $d$ \cite{gosch, mall1, mall4, malrap3}). In a (co)homological-algebraic sense, 
\emph{the curvature of an algebraic connection measures the `obstruction' to or the `deviation' from the nilpotency of the connection (:differential)} \cite{malrap1, malrap2, malrap3}.}

As a matter of fact, it is rather straightforward to see that, for
$\modl$ a vector sheaf, $\curv(\conn)$ {\em is functorially defined as an $\struc$-morphism of
$\struc$-modules}, in the following sense:

\begin{equation}\label{eq19}
\begin{array}{c}
\curv\in{\mathrm{Hom}}_{\struc}(\modl
,\Omg^{2}(\modl))=\Hom_{\struc}(\modl,\Omg^{2}(\modl))(X)\cr
\Omg^{2}({\modl}nd\modl)(X)=Z^{0}(\gauge,\Omg^{2}({\modl}nd\modl))
\end{array}
\end{equation}

\noindent where, as usual, $\gauge_{i}=\{ U_{\alpha}\}_{\alpha\in I}$ is an open
cover of the base topological space $X$ and $Z^{0}(\gauge,\Omg^{2}(\mathcal{E}nd\modl))$ the
$\struc(U)$-module of $0$-{\em cocycles} of
$\Omg^{2}(\modl nd\modl)$ relative to the
$\gauge_{i}$-covering of $X$.\footnote{One may wish to recall again
that, for a vector sheaf $\modl$ like the one involved in
(\ref{eq19}) above: ${\modl}nd\modl\equiv{\mathcal{H}}om_{\struc}(\modl
,\modl)\cong\modl\otimes_{\struc}\modl^{*}=\modl^{*}\otimes_{\struc}\modl$.}

\subsubsection{The Non-Tensorial and the Tensorial Character of $\conn$ and $\curv$, respectively, under Local `Gauge-Coordinate' Transformations}

(\underline{\bf Note:} In the following presentation and discussion, we are not going to specify what ADG-connections and their curvatures we are talking about. The reader can assume that the connections are either Einstein-Lorentzian or Yang-Mills, in the sense that the arguments below apply {\it mutatis mutandis} to both.)

In this subsection, we recall from \cite{mall1, malrap3, mall4} a very subtle and important for our arguments in the sequel ADG-theoretic result, which may be distilled down to the following two statements:

\begin{quotation}

\noindent $\bullet$ {\em The ADG-theoretic connection $\conn$ is only a $\cons$-linear sheaf morphism (hence not an $\otimes_{\struc}$-tensor)};\footnote{Where $\cons$ is just \emph{the constant sheaf of $\com$-numbers} \cite{mall1, malrap3, mall4}.} while;

\noindent $\bullet$ {\em The ADG-theoretic curvature $\curv$ is a full $\struc$-structure sheaf morphism (hence a pure $\otimes_{\struc}$-tensor)}.\footnote{With $\otimes_{\struc}$ the {\em homological algebraic (:categorical) tensor product functor}. In the sequel, we will return to explain and discuss in more detail the paramount importance of $\otimes_{\struc}$ and its adjoint functor $\mathrm{Hom}$ \cite{maclane1, maclane2, macmo} for the {\em dynamical functoriality} of our (finitary) ADG-perspective on gravity and gauge theories, and its cogent physical interpretation in the quantum deep.}

\end{quotation}

\noindent Two equivalent statements to the ones above, which the theoretical physicist/mathematician who is familiar with the usual differential geometry of gauge theory, which employs smooth fiber bundles over a $\smooth$-smooth differential spacetime manifold $M$ can straightforwardly understand \cite{gosch}, are the following:

\begin{quotation}

\noindent $\bullet$ {\em $\conn$ does transforms inhomogeneously (:non-tensorially or affinely) under local gauge-coordinate transformations}; while;

\noindent $\bullet$ {\em $\curv$ does indeed transform homogeneously (:tensorially) under local gauge-coordinate transformations}.

\end{quotation}

\noindent To explicate in detail what the above statements mean, let us recall briefly from \cite{mall1, malrap3, mall4} how the ADG-connection $\conn$ and the ADG-curvature $\curv$ behave (:transform) respectively under local general coordinate-gauge changes.

\subsubsection{Local Gauge Transformation of $\conn$} 

Let $\modl$ be a differential $\struc$-module (:an ADG-theoretic vector sheaf) of rank
$n$. Let $e^{U}\equiv\{ U;~ e_{i=1\cdots n}\}$ and $f^{V}\equiv\{
V;~ f_{i=1\cdots n}\}$ be local gauges\footnote{A general gauge-coordinate $n$-\emph{frame} (or $n$-\emph{bein}).} of $\modl$ over the open
set gauges $U$ and $V$ of $X$\footnote{$U,V\in\gauge_{i}$, with $\gauge_{i}$ an open covering of the underlying topological space $X$ of the $\cons$-algebraized space $(X,\struc)$, as assumed throughout this paper.} which, in turn, we assume have non-empty
intersection ($U\cap V\not= \emptyset$). Let us also denote by $g\equiv(g_{ij})$ the
following {\em change of local gauge matrix}:

\begin{equation}\label{eq20}
f_{j}=\sum_{i=1}^{n}g_{ij}e_{i}
\end{equation}

\noindent which, plainly, is a local ({\it i.e.}, relative to $U\cap
V$) section of the `natural' structure group sheaf $\gl(n,\struc)$
of $\modl$\footnote{As noted earlier, one may recognise
$\gl(n,\struc)$ above as the local version of the automorphism principal
group sheaf $\aut_{\struc}\modl$ of $\modl$. The adjective `local' here
pertains to the fact mentioned earlier that ADG assumes that
$\modl$ is locally isomorphic to $\struc^{n}$.}---that is to say,
$g_{ij}\in\mathrm{GL}(n,\struc(U\cap V))=\gl(n,\struc)(U\cap V)$.

Without going into the details of the derivation, which can be
found in \cite{mall1, mall2, mall4}, we note that under such a local gauge
transformation $g$, the gauge potential part $\aconn$
of $\conn$ in (\ref{eq13}) transforms as follows:

\begin{equation}\label{eq21}
\aconn^{'}=g^{-1}\aconn g+g^{-1}\partial g
\end{equation}

\noindent a way we are familiar with from the usual differential
geometry of the smooth fiber bundles of gauge theories \cite{gosch}.  For
completeness, it must be noted here that, in (\ref{eq21}),
$\aconn\equiv(\aconn_{ij})\in
M_{n}(\Omg^{1}(U))=M_{n}(\Omg^{1})(U)$ and
$\aconn^{'}\equiv(\aconn^{'}_{ij})\in
M_{n}(\Omg^{1}(V))=M_{n}(\Omg^{1})(V)$. 

\begin{quotation}

\noindent The transformation of
$\aconn$ under local gauge changes is called {\em
inhomogeneous}, {\em non-tensorial} or {\em affine} in the usual gauge-theoretic parlance \cite{gosch} precisely
because of the (additional to the homogeneous) term $g^{-1}\partial g$. 

\end{quotation}

\subsubsection{Local Gauge Transformation of $\curv$} 

On the other hand, we read directly from \cite{mall1, malrap3, mall4} that under similar local gauge-coordinate changes, the curvature $\curv(\conn)$ of the ADG-connection $\conn$ transforms {\em purely homogeneously} or {\em tensorially}, as follows:

\noindent To that end, let again $g\equiv g_{ij}\in\gl(n,\struc)(U\cap V)$ be the
change-of-gauge matrix we considered in (\ref{eq20}) in connection
with the transformation law of gauge potentials $\aconn_{ij}$. Again, without
going into the technical details of the derivation, we bring forth from
\cite{mall1, malrap3, mall4} the following {\em local transformation law of gauge
field strengths}:

\begin{equation}\label{eq22}
\begin{array}{c}
\mathrm{for~a~local~frame~change :~}
e^{U}\stackrel{g}{\mapto}e^{V} (U,V\in\gauge{i}~\mathrm{covering}~X),\cr
\mathrm{the~curvature~transforms~as:}~\curv\stackrel{g}\mapto\curv^{'}=g^{-1}\curv
g
\end{array}
\end{equation}

\noindent the form of which we are familiar with from the usual differential
geometric ({\it i.e.}, smooth fiber bundle-theoretic) treatment of
gauge theories \cite{gosch}. For completeness, we remind ourselves here that,
in (\ref{eq21}) above, $\curv^{U\cap V}\equiv(\curv^{U\cap V}_{ij})\in
M_{n}(\Omg^{2}(U\cap V))$---an $(n\times n)$-matrix of sections of
local $2$-forms in $\Omg^{2}$. 

\begin{quotation}

\noindent The transformation of $\curv$ under local gauge-coordinate
changes is called {\em homogeneous}, {\em tensorial} or {\em covariant} in the
usual smooth fiber bundle gauge-theoretic parlance \cite{gosch}. 

\end{quotation}

As a last important observation before we move on to explicate the $\struc$-invariant and its associated $\otimes_{\struc}$-functorial character of the dynamical equations of motion from Einstein gravity and free Yang-Mills gauge theories, we note:

\subsubsection{ADG-curvature spaces and ADG-curvature field categories} 

As we increase by a notch the level of abstraction and generality, from \cite{mall1, malrap3, mall4} we note the definition of \emph{ADG-curvature spaces} as the following quintuples:

\begin{equation}\label{eq23}
(\struc ,\conn, \Omg^{1},\conn^{2} ,\Omg^{2})
\end{equation}

\noindent consisting of $\struc$-modules and $\cons$-linear morphisms between them, which, in turn, by the very definition of the ADG-curvature field in (\ref{eq17}), reduce to the following duet representing \emph{the ADG-curvature fields}:

\begin{equation}\label{eq24}
\mathcal{R}:=(\struc , \curv(\conn))
\end{equation}

\subsubsection{Third encounter with `Dynamical' Functoriality: three ADG-curvature field functor categories} 

In much the same way that we defined earlier the three functor categories of ADG-connection fields: the {\em Maxwell} $\mathcal{T}_{Max}$, the {\em Yang-Mills} $\mathcal{T}_{YM}$, and the {\em Einstein} category $\mathcal{T}_{Einst}$ earlier, we can similarly define here:

\begin{quotation}

\noindent {\em Three ADG-curvature field functor categories}: $\mathcal{C}_{Max}$, $\mathcal{C}_{YM}$ and $\mathcal{C}_{Einst}$, whose objects are ADG-curvature fields as in (\ref{eq24}), and whose arrows are \emph{natural transformation} type of correspondences between their $\otimes_{\struc}$-functorial objects.\footnote{We are going to return to this important definition shortly, when we explicate the $\struc$-invariant and $\otimes_{\struc}$-functorial Einstein and Yang-Mills ADG-theoretic local gauge dynamical laws of motion. Of special interest and semantic importance will be the pair of adjoint functors $\mathrm{Hom}$-$\otimes_{\struc}$, which will be seen to be a \emph{geometric morphism/natural transformation type of correspondence} between the corresponding ADG-categories of connection fields $\conn$ and their curvatures $\curv(\conn)$: for gravity, for instance, the mapping: $\xymatrix{
\mathcal{T}_{Einst} \ar[r]^{\mathrm{Hom}-\otimes_{\struc}} & \mathcal{C}_{Einst}\\}$ will be seen to be such a {\em geometric morphism} of huge physical significance for the cogent physical semantics of our ADG-theoretic perspective on gravity and gauge theories as Mallios had originally envisaged.}

\end{quotation}

\subsubsection{Lagrangean Action Derivation of Vacuum Einstein Gravity and Free Yang-Mills Theories} 

Now that we have recalled the essential characteristics and local gauge transformation behaviour of the affine ADG-connections and their curvature $\otimes_{\struc}$-tensors, we note that the ADG-theoretic versions of \emph{the dynamical free Yang-Mills and vacuum Einstein equations} both derive from (the variation of) respective Yang-Mills ($\ym$) and Einstein-Hilbert ($\eh$) \emph{Lagrangean action functionals}, as follows:\footnote{For the equations above, see \cite{mall1, mall4} for technical definitions and details.}

\begin{equation}\label{eq25}
\begin{array}{c}
\eh_{\modl}(\conn)=\int_{X}tr(\curv_{Ric}(\conn))\;\stackrel{\delta\aconn}{\mapto}\;\curv_{Einst}(\modl)=0 \cr
{}\cr
\ym_{\modl}(\conn)=\frac{1}{2}\int_{X}tr(\curv_{YM}\wedge\star\curv_{YM})\;\stackrel{\delta\aconn}{\mapto}\;\Delta_{\modl
nd\modl}^{2}(\curv_{YM})=0
\end{array}
\end{equation}

\noindent in which we read from \cite{mall1, malrap3, mall4} that $\curv_{Einst}$ is the {\em Ricci Scalar}\footnote{For expository completeness, we briefly recall from \cite{malrap3} that given a (real) Lorentzian vector sheaf $(\modl ,\rho)$ of rank $n$
equipped with a non-flat Lorentzian $\rho$-metric $\struc$-connection $\conn$, one can define the following {\em Ricci curvature operator} $\curv_{Ric}$ relative to a local gauge $U\in\gauge_{i}$ of $\modl$ : $\curv_{Ric}(\, .\, ,\,\! s)t\in (\modl nd\modl)(U)=M_{n}(\struc(U))$, for local sections $s$ and $t$ of $\modl$ in
$\modl(U)=\struc^{n}(U)=\struc(U)^{n}$. Thus, the Ricci curvature here $\curv_{Ric}$ is an $\modl
nd\modl$-valued operator, a {\em curvature endomorphism of} $\modl$. Moreover, {\em since $\curv_{Ric}$ is matrix-valued, one can take its trace}, thus define the following {\em Ricci scalar
curvature operator} $\curv_{Einst}:=tr(\curv_{Ric}(\, .\, ,\,\! s)t)$, which, plainly, is $\struc(U)$-valued.} of the ADG-theoretic Einstein-Lorentzian metric connection field $\conn$, while $\curv_{YM}$ is the {\em Yang-Mills gauge field strength} of the homonymous ADG-theoretic Yang-Mills connection field $\conn$.

\subsection{Miscellaneous Remarks on `Dynamical Functoriality' and $\struc$-Invariance-cum-Covariance of the ADG-Field Autodynamics: A Unified, Pure Gauge, Smooth Base Spacetime Manifoldless and Finitistic ADG-Theoretic Quantum Field `Solipsism' of the 3rd Kind}

In this subsection we make eight conceptual and technical remarks on the ADG-theoretic perspective on the Einstein and Yang-Mills field dynamics in (\ref{eq25}) above. We itemise our remarks as follows:

\subsubsection{ADG-Kinematics: The Affine Space of $\struc$-Connections} 

From the vacuum Einstein and the free Yang-Mills dynamical equations \`a-la ADG in (\ref{eq25}) above, it follows that the sole dynamical variable in our ADG-theoretic perspective on gravity and gauge theories is the local $\struc$-connection $\conn$. That is, as equation (\ref{eq25}) depicts above, the dynamical equations derive from the variation ($\delta\aconn$) of the Einstein-Hilbert and Yang-Mills Lagrangian action functionals with respect to the local gauge potential part ($\aconn$) of the gravitational and Yang-Mills $\struc$-connections $\conn$ on their respective vector sheaves $\modl$.

Thus, as emphasised in \cite{malrap3}, the sole dynamical variable in our ADG-theoresis of vacuum Einstein gravity and free Yang-Mills theories is the ADG-theoretic Einstein connection field pairs $\mathcal{F}_{Einst}=(\modl ,\conn)$ and $\mathcal{F}_{YM}=(\modl ,\conn)$ defined by (\ref{eq14}) earlier, within their respective categories $\mathcal{T}_{Einst}$ and $\mathcal{T}_{YM}$.

As also highlighted in \cite{malrap3}, it follows that {\em the generalised `kinematical' space of the theory is the affine space $\sconn_{\struc}(\modl)$ of $\struc$-connections $\aconn$ on $\modl$}. Moreover, since the Lagrangians involved in (\ref{eq25}) are invariant under the group sheaf $\mathcal{G}(\modl)=\aut_{\struc}\modl$ of local automorphisms (:local gauge transformations) of $\modl$, 

\begin{quotation}
\noindent {\em the relevant kinematical space is the moduli space $\sconn /\mathcal{G}=\sconn_{\struc}(\modl) /\aut_{\struc}\modl$ of gauge equivalent $\struc$-connections $\aconn$ on $\modl$}.\footnote{Recall from \cite{malrap3} that $\mathcal{G}(\modl)|_{U}=\aut_{\struc}\modl |_{U}=(\mathrm{Aut}_{\struc}\modl)(U):=\Gamma(U,\aut_{\struc}\modl)\equiv M^{n}_{U}(\struc)$. Thus, $\sconn /\mathcal{G}=\sconn_{\struc}(\modl) /\aut_{\struc}\modl$ is the so-called $\mathcal{G}${\em -orbit space} as the structure gauge group sheaf $\aut_{\struc}\modl(\struc)$ cuts through the affine space $\sconn_{\struc}(\modl)$, carving out `paths' or `orbits' of gauge equivalent connections in the process, which, in turn, leave the corresponding curvature Lagrangians in (\ref{eq25}) invariant under (local) gauge transformations.}
\end{quotation}

\noindent Hence the integration sign in the dynamical action functionals in (\ref{eq25}), which supposedly extends over the base topological space $X$, in effect extends over the moduli space $\sconn /\mathcal{G}=\sconn_{\struc}(\modl) /\aut_{\struc}\modl$ of gauge equivalent connections.\footnote{From \cite{malrap3, mall14, mall13, mall4} we read that {\em a suitably defined ADG-theoretic Radon-type of $\struc$-linear continuous integration measure $d\mu$ on a suitably topologised} $\sconn /\mathcal{G}$ ($d\mu :\; \sconn_{\struc}(\modl) /\aut_{\struc}\modl\longrightarrow\struc$) {\em is expected to render rigorous the dynamical action integrals in} (\ref{eq25}).} 

\subsubsection{$\struc$-Invariance and `Dynamical' Functoriality} 

We noted earlier a fundamental difference in ADG between an $\struc$-connection $\conn$ and its curvature $\curv(\conn)$, namely that,

\begin{quotation}

\noindent {\em The curvature $\curv(\conn)$ is an $\otimes_{\struc}$-tensor, while the connection $\conn$ itself is not}.

\end{quotation}

\noindent In other words,

\begin{quotation}

\noindent {\em The curvature $\curv(\conn)$ respects our (algebras) of generalised measurements in $\struc$, while the connection $\conn$ itself does not---it `eludes' them}.

\end{quotation}

\noindent In still equivalent parlance,

\begin{quotation}

\noindent {\em The curvature $\curv(\conn)$ is a `geometrical object', while the connection $\conn$ itself is an `algebraic object'}.\footnote{We will make this statement mathematically much more precise and rigorous in the sequel when we discuss the fundamental $\otimes_{\struc}-\mathrm{Hom}$-adjunction `{\em geometric morphism equivalence}' between the ADG Einstein and/or Yang-Mills connection field and curvature field functor categories $\mathcal{T}_{Einst/YM}$ and $\mathcal{C}_{Einst/YM}$, respectively.}

\end{quotation}

\noindent Yet, 

\begin{quotation}

\noindent {\em the dynamics on $\modl$, generated by the connection field $\conn$ acting on $\modl$ and expressed as a differential equation on it as in (\ref{eq25}), is derived from an action principle involving the curvature of the connection}.

\end{quotation}

\noindent As such,

\begin{quotation}

\noindent {\em the dynamical equations of motion on $\modl$, which are derived from an action principle involving the curvature of the connection, is gauge invariant, hence our free gauge choices of generalised coordinate measurements in $\struc$ respect the dynamics, and the physical laws are independent of our measurements in $\struc$}.

\end{quotation}

\noindent Thus, {\it in toto},

\begin{quotation}

\noindent {\em The physical laws are $\struc$-invariant}.

\end{quotation}

\noindent Which brings us to a fundamental observation, in connection with {\em Utiyama's Theorem}, that we read directly from \cite{mall13}.

\subsubsection{The Algebra-Geometry Duality: The $\mathrm{Hom}_{\struc}-\otimes_{\struc}$ Functorial Adjunction between the ADG-Field and Curvature Categories} 

Below, we quote Mallios {\it verbatim} from \cite{mall13}:

\begin{quotation}

\noindent ``...{\em Utiyama's theorem}, relates/characterizes the `{\em $\struc$-invariance}' of what we may call {\em $\struc$-connection Lagrangian through that one of} the corresponding {\em curvature Lagrangian}. So {\em the} aforementioned {\em two notions} (of `{\em Lagrangians}') are, in effect (physically) {\em equivalent}, through/due to the `$\struc$-{\em invariance}'...''\footnote{Throughout this quotation we have been faithful to the {\em emphasis} placed by Mallios on certain key words in the original paper \cite{mall13}.}

\end{quotation}

\noindent which leads us to Mallios's telling remarks of what he calls {\em The Fundamental (Physical) Adjunction}:\footnote{Again quoting Mallios {\it verbatim} from \cite{mall13}.}

\begin{quotation}

\noindent ``...{\bf The Fundamental (Physical) Adjunction:}  Thus, the basic Homological (:categorical) $Hom-\otimes$ {\em adjunction},  corresponds, {\em within the context of ADG}, to the {\em fundamental physical adjunction}, effectuated by the following `{\em adjoint pair of functors}':

$$\struc-connection\;(:field,\; `potential')\leftrightarrows\;curvature\;(:`field\; strength')$$

\noindent The above can actually be perceived, as describing the whole {\em function of a physical law}, hence, in fact, {\em of the Nature herself}...''\footnote{Again,throughout the quotation above we have been faithful to the {\em emphasis} placed by Mallios on certain key words in the original paper \cite{mall13}.}

\end{quotation}

\noindent And Mallios concludes Section 2 of the paper \cite{mall13} with the following intuitively telling paragraph:

\begin{quotation}

\noindent ``...On the other hand, the {\em connecting function} of a given {\em adjunction}, is in effect a {\em natural transformation of functors}. Consequently, the latter should still preserve `{\em $\struc$-invariance}' of the adjunction, with respect to any {\em `$\struc$-invariant function', referring to any one} of the two associated functors through the adjunction: One gets at it, just, based on the very definitions\footnote{Given before in the paper \cite{mall13}.} and on the {\em `functorial nature' of ADG}...''\footnote{Once again, throughout the quotation above, we have been faithful to the {\em emphasis} placed by the author on certain key words in the original text \cite{mall13}.}

\end{quotation}

\noindent Now, in view of our presentation and arguments in the present paper, we are in a position to distill and further mathematically formalise and explicate Mallios's remarks above on {\em functoriality, adjunction} and {\em natural transformation of functors}.

\subsubsection{Fourth Encounter with `Dynamical' Functoriality: Mallios's Fundamental (Physical) Adjunction Explicated and Interpreted} 

The {\em categorical adjunction between the connection $\conn$ and its curvature $\curv(\conn)$}

\begin{equation}\label{eq26}
\struc-connection\rightleftarrows\struc-curvature
\end{equation}

\noindent that Mallios emphasises in the excerpt from \cite{mall13} in connection with Utiyama's Theorem displayed above, can now be cast in a mathematically rigorous and precise {\em functorial form}, as {\em a functorial correspondence between the respective categories of Einstein (or Yang-Mills) connection fields} $\mathcal{T}_{Einst}$ (or $\mathcal{T}_{YM}$) {\em and their corresponding categories of Einstein (or Yang-Mills) curvature field strengths} $\mathcal{C}_{Einst}$ (or $\mathcal{C}_{YM}$), as follows:

\begin{equation}\label{eq27}
\begin{array}{c}
\xymatrix{
\mathcal{T}_{Einst} \ar@<1ex>[r]^{\otimes_{\struc}}
& \mathcal{C}_{Einst} \ar@<1ex>[l]^{\mathrm{Hom}_{\struc}} }\cr
{}\cr
\xymatrix{
\mathcal{T}_{YM} \ar@<1ex>[r]^{\otimes_{\struc}}
& \mathcal{C}_{YM} \ar@<1ex>[l]^{\mathrm{Hom}_{\struc}} }
\end{array}
\end{equation}

\noindent with the pair of `opposite direction maps' $(\otimes_{\struc},\mathrm{Hom}_{\struc})$ above corresponding to {\em the fundamental homological algebraic  adjunction} that Mallios alludes to.

More technically speaking, $\otimes_{\struc}$ is {\em the homological (left-adjoint) tensor product functor}\footnote{In the category of $\struc$-modules that the vector sheaves $\modl$ of ADG belong.} and $\mathrm{Hom}_{\struc}$ {\em is its right-ajoint functor} \cite{macmo}. When paired together, the pair:

\begin{equation}\label{eq28}
\mathcal{G}\mathcal{M}_{\struc}:=(\otimes_{\struc},\mathrm{Hom}_{\struc})
\end{equation}

\noindent constitutes an instance of what is commonly known in category theory as a {\em geometric morphism} \cite{macmo}.\footnote{It is instructive here to give the definition of a general {\em geometric morphism} directly from \cite{macmo}, as it originally arose in category theory. With every continuous map $f$ between two topological spaces $X$ and $Y$: $f:\, X\rightarrow Y$, there is {\em a pair of adjoint functors} $(f^{*},f_{*})$: $\xymatrix{
Sh(X) \ar@<1ex>[r]^{f_{*}}
& Sh(Y) \ar@<1ex>[l]^{f^{*}} }$ ($f_{*}$ is coined the {\em push-out} and $f^{*}$ is coined the {\em pull-back}) between the categories of sheaves (of structureless sets) $Sh(X)$ and $Sh(Y)$ over $X$ and $Y$, respectively.}

At the same time, the fact that the pair $(\otimes_{\struc},\mathrm{Hom}_{\struc})$ indeed constitutes a categorical adjunction, derives directly from the way $\otimes_{\struc}$ and $\mathrm{Hom}_{\struc}$ act on the corresponding categories. Again, in the general case, we read from \cite{macmo} that $-\otimes X$ is the {\em left-adjoint} (functor) and $\mathrm{Hom}(X,-)$ the {\em right-ajoint} (functor), because they act as a pair of maps as follows:

\begin{equation}\label{eq29}
\mathrm{Hom}(Y\otimes X, Z)\simeq \mathrm{Hom}(Y,\mathrm{Hom}(X,Z))
\end{equation}

\noindent Thus, in view of the general definition of {\em the action of an adjunction} as in (\ref{eq29}) above, we are now in a position to see directly that,  indeed: 

\begin{quotation}
\noindent {\em In ADG, the curvature $\curv(\conn)$ of a connection is the $\otimes_{\struc}$-morph (:image) of its connection.}
\end{quotation}

\noindent which we can verify directly from the curvature's definition in terms of the action of the $\otimes_{\struc}$ and $\mathrm{Hom}_{\struc}$ functors in (\ref{eq17}) and (\ref{eq19}) earlier:

\begin{equation}\label{eq30}
\begin{array}{c}
\curv :\;\modl\stackrel{\mathrm{Hom}_{\struc}}{\longrightarrow}\Omg^{2}(\modl)\equiv\modl\otimes_{\struc} \Omg^{2}\cr
\curv\in{\mathrm{Hom}}_{\struc}(\modl
,\Omg^{2}(\modl))=\Hom_{\struc}(\modl,\Omg^{2}(\modl))(X)
\end{array}
\end{equation}

\noindent Recalling again from footnote 26 that, for any ADG-theoretic vector sheaf $\modl$ like the one involved in
(\ref{eq30}) above: ${\modl}nd\modl\equiv{\mathcal{H}}om_{\struc}(\modl
,\modl)\cong\modl\otimes_{\struc}\modl^{*}=\modl^{*}\otimes_{\struc}\modl$.

We can distill all the above discussion and express the {\em geometric morphism functorial equivalence} between the connection-field and curvature-field Einstein and/or Yang-Mills {\em functor categories} $\mathcal{T}_{Einst/YM}$ and $\mathcal{C}_{Einst/YM}$ depicted in equations (\ref{eq27}) and (\ref{eq28}) above, as follows:

\begin{equation}\label{eq31}
\begin{array}{c}
\mathcal{T}_{Einst/YM}\stackrel{\mathcal{GM}}{\longleftrightarrow}\mathcal{C}_{Einst/YM}\cr
{}\cr
\xymatrix{
\mathcal{T}_{Einst/YM} \ar@<1ex>[r]^{\otimes_{\struc}}
& \mathcal{C}_{Einst/YM} \ar@<1ex>[l]^{\mathrm{Hom}_{\struc}} }\cr
{}\cr
\conn_{Einst/YM}\stackrel{\otimes_{\struc}}{\longrightarrow}\curv_{Einst/YM}
\end{array}
\end{equation}

\noindent with the first two lines in (\ref{eq31}) above, reading:

\begin{quotation}
\noindent {\em The curvature (Einstein and/or Yang-Mills field categories) are the geometric morphs (:images) of the corresponding connection (field categories)}.
\end{quotation}

\noindent while the map in the third line of (\ref{eq31}) above can be interpreted as stipulating that:

\begin{quotation}
\noindent {\em The (Einstein and/or Yang-Mills) curvature field is the $\otimes_{\struc}$-image of its connection field}.
\end{quotation}

\noindent which in turn vindicates what we established earlier, namely that:

\begin{quotation}
\noindent {\em The curvature is a geometrical, $\struc$-tensorial object (:an $\otimes_{\struc}$-tensor or an $\struc$-invariant morphism), while its connection is not}.
\end{quotation}

\noindent This gives a {\it raison d'\^{e}tre} and vindicates the epithet \underline{\em geometric} in the {\em geometric morphism} denomination of the pair of maps $\mathcal{G}\mathcal{M}_{\struc}:=(\otimes_{\struc},\mathrm{Hom}_{\struc})$ above.

\subsection{$\mathcal{G}\mathcal{M}$-Dynamical Functoriality, Mallios's $\struc$-Invariance and Gauge Invariance of ADG-GT}

Now that we have explicated the subtle and technical sense in which a {\em geometric morphism adjunction} $\mathcal{G}\mathcal{M}$ links the ADG-theoretic Einstein/Yang-Mills connection field category $\mathcal{T}_{Einst/YM}$ with its ADG-theoretic Einstein/Yang-Mills curvature field `counterpart-equivalent' category $\mathcal{T}_{Einst/YM}$, we are in a position to further support {\em Mallios's remarks on $\struc$-invariance in connection with gauge invariance} in \cite{mall13}.

To this end, we quote {\it verbatim} from Section 3 of \cite{mall13} the displayed paragraph before Theorem 3.1:

\begin{quotation}

\noindent ``...Therefore, one thus realizes that, {\em the fundamental adjunction\footnote{Our geometric morphism $\mathcal{G}\mathcal{M}$ in equations (\ref{eq27}) and (\ref{eq28}) earlier.}, preserves the $\struc$-invariance, for any $\struc$-invariant function,\footnote{Especially, for the $\struc$-invariant Einstein and Yang-Mills action functionals in (\ref{eq25}) above, which, as we saw earlier, are $\struc$-valued functionals defined on the moduli space $\sconn_{\struc} /\mathcal{G}=\sconn_{\struc}(\modl) /\aut_{\struc}\modl$ of gauge equivalent $\struc$-connections on $\modl$.} pertaining to the two basic functors appearing in the aforesaid adjunction...''\footnote{That is, the homological $\mathrm{Hom}_{\struc}$ and $\otimes_{\struc}$ adjoint functors constituting $\mathcal{G}\mathcal{M}_{\struc}$ in equations (\ref{eq27}) and (\ref{eq28}) earlier.}}

\end{quotation}

\noindent which in turn leads to the following central result (:Theorem 3.1) in \cite{mall13}, coined {\em Utiyama's Principle} therein:\footnote{Again, quoted exactly as it appears in \cite{mall13}.}

\begin{quotation}

\noindent ``...{\bf Theorem 3.1} {\em Any `gauge invariance' of an appropriate `Lagrangian' for ($\struc$-)connections is equivalent to a similar invariance of the corresponding Lagrangian for the associated curvature with the ($\struc$-)connection at issue...}''

\end{quotation}

\noindent Thus, we are now in a position to distill and re-express the deep relation between our notion of {\em `Dynamical' Functoriality}, with Mallios's {\em $\struc$-invariance and the structure group $\aut_{\struc}\modl$-gauge invariance (of the dynamical action functionals) of our ADG-theoresis on Vacuum Einstein Gravity and Free Yang-Mills Theories} in (\ref{eq25}), as follows:

\begin{quotation}
\noindent {\bf Fundamental Theorem of ADG-Gauge Theory.} {\em The `dynamical' geometric morphism $\mathcal{G}\mathcal{M}_{\struc}$ preserves $\struc$-invariance and entails the structure group sheaf $\mathcal{G}\equiv\aut_{\struc}\modl$-invariance of the ADG-theoretic dynamical equations of motion for Vacuum Einstein Gravity and Free Yang-Mills Theory}.
\end{quotation}

In view of our physical interpretation of $\struc$ earlier in this paper and throughout our work on ADG-GT \cite{malrap1, malrap2, malrap3, mall4, rap5, rap7, rap11, rap13, rap14} as {\em the algebra (sheaf) of our generalised coordinate measurements},\footnote{That is, our localised and gauged $\struc$-valued measurements based on an open cover $\{ U\}=\gauge_{i}$ of local open sets $U$ of the base topological space $X$ employed by the theory.} an important Corollary to the Fundamental Theorem above {\em goes the other way around}, as follows:

\begin{quotation}
\noindent {\bf Corollary to the Fundamental Theorem of ADG-Gauge Theory.} {\em The algebra (sheaf) of our generalised coordinate measurements $\struc$ respects ({\it i.e.} it is `non-perturbing' and it leaves invariant) the functorial ADG-theoretic gauge field dynamics for Vacuum Einstein Gravity and Free Yang-Mills theories, hence, in return, it entails and almost `mandates' that the (local) Relativity Group (sheaf) of the theory is} $\mathcal{G}\equiv\aut_{\struc}\modl$.\footnote{That the local relativity group in ADG-GT is `naturally' $\aut_{\struc}\modl$ (:see subsection next) has been amply expounded in \cite{malrap1, malrap2, malrap3, mall4, mall14, mall15, rap5, rap7, rap11, rap13, rap14}.} Thus, $\struc$-invariance, via the `dynamical' geometric morphism $\mathcal{G}\mathcal{M}_{\struc}$, which in turn entails the $\mathcal{G}=\aut_{\struc}\modl$-gauge invariance of the ADG-field dynamics (expressed via the geometric morph of the connection field---the field's curvature), corresponds to a dynamical version of the {\em Kleinian conception of geometry}.\footnote{According to Felix Klein, `{\em the geometry of an object is all the transformations of it that leave it invariant}' \cite{klein}.}
\end{quotation}

\noindent The discussion above brings us to an important, yet subtle, technical and interpretational matter of ADG-GT.

\subsubsection{Note on the Natural Transformation character of the Geometric Morphism $\mathcal{GM}_{\struc}$: the `Naturality' of the Functorial Dynamics of ADG-GT} 

Since, as we alluded to numerous times throughout this paper, the ADG-theoretic connection and curvature field categories are {\em functor categories} \cite{mall13, mall14, mall15, mall4}, the two adjoint functors constituting the geometric morphism $\mathcal{G}\mathcal{M}_{\struc}:=(\otimes_{\struc},\mathrm{Hom}_{\struc})$ above are examples of {\em natural transformations} \cite{macmo}.\footnote{In a nutshell, and quite heuristically, {\em a natural transformation $\mathcal{N}$ connects or maps one particular functor $\mathcal{F}:\, A\rightarrow B$ to another particular functor $\mathcal{G}:\, A\rightarrow B$ between two categories $A$ and $B$}. At the same time, $\mathcal{N}$ {\em does not need to apply to every functor in some category of functors} \cite{macmo}.}

This, too, was prophetically anticipated by Mallios in \cite{mall13}:\footnote{Excerpt from quotation earlier.}

\begin{quotation}
\noindent ``...On the other hand, the {\em connecting function} of a given {\em adjunction}, is in effect \underline{\em a natural transformation of functors}. Consequently, the latter should still preserve `{\em $\struc$-invariance}' of the adjunction, with respect to any {\em `$\struc$-invariant function', referring to any one} of the two associated functors through the adjunction...''
\end{quotation}

\noindent The fitting physico-mathematical `pun' here is that: 

\begin{quotation}

\noindent {\em The $\struc$-invariant and, in extenso, $\aut_{\struc}\modl$-invariant functorial gauge field dynamical changes in ADG-GT are, categorically-cum-physically speaking, \underline{Natural Transformations} of the `dynamically equivalent' gauge connection and curvature fields involved therein, via the Natural Transformation Geometric Morphism $\mathcal{G}\mathcal{M}_{\struc}:=(\otimes_{\struc},\mathrm{Hom}_{\struc})$ that interlinks them}.

\end{quotation}

\section{Intermezzo: Lateral Technical and Philosophical Offshoots and Repercussions of Functoriality}

In this intermediate section, we give very brief accounts and we express them in the form of `{\em Aphorisms}', borrowed from previous works, of various technical, conceptual and interpretational-cum-philosophical corollaries and didactics that follow, in one way or another, from both the 'kinematical' and `dynamical' functoriality of our ADG-theoretic perspective on Finitary Vacuum Einstein-Lorentzian Gravity and Free Yang-Mills Theories, as expounded above.

\subsection{Third Gauge Auto-Gravitodynamics from Background Spacetime Manifoldlessness: Gauge Field Solipsism} 

The quintessential feature of ADG, especially {\it vis-\`a-vis} its novel conceptual import and potential technical applications to Quantum Gravity research \cite{mall1, mall2, mall4, mall5, mall6, mall7, mall9, mall10, mall11, mall12, mall13, mall14, mall15, malrap1, malrap2, malrap3, malrap4, malros1, malros2, malros3, malzaf2, malzaf1, rap1, rap2, rap5, rap7, rap11, rap13, rap14, rapzap1, rapzap2, zaf1, zaf2}, is arguably the following:

\begin{quotation}

\noindent {\bf  \underline{Aphorism 1:} Background Spacetime Manifoldlessness.} {\em Mallios's Abstract Differential Geometry is a purely homological-algebraic (:sheaf and category-theoretic) way of doing and applying `Differential Calculus', with all its technical and conceptual panoply, to many current research fronts in Theoretical and Mathematical Physics such as Quantum Gauge Theories of Matter and Quantum Gravity, but in the manifest absence of a background geometrical $\smooth$-smooth base (spacetime) manifold.}

\end{quotation}

\noindent The deep and wide spectrum of potential import of such {\em background differential spacetime manifoldlessness}, especially in Quantum Gauge Theory and Quantum Gravity research, has been expounded in detail over the last two decades in numerous works \cite{mall5, mall3, mall6, mall9, mall14, mall15, mall4, malrap1, malrap2, malrap3, rap5, rap7, rap11, rap13, rap14, malzaf2, malzaf1, zaf1, zaf2}.

One important feature of such {\em a background spacetime manifold independence} is that:

\begin{quotation}

\noindent {\bf \underline{Aphorism 2:} Dynamical Connection Gauge Field Solipsism.} ADG enables us to formulate Vacuum Einstein Gravity and Free Yang-Mills theories as {\em pure gauge theories of the third kind},\footnote{We recall directly from \cite{malrap3, rap13, rap14} that {\em Gauge Theory of the First Kind} is Hermann Weyl's original Global ($U(1)$) Scale Theory of the Electromagnetic Field \cite{weyl}; {\em Gauge Theory of the Second Kind} pertains to the usual current abelian and non-abelian Yang-Mills gauge theories of matter that are localised and gauged based on an externally prescribed and fixed $\smooth$-smooth base spacetime continuum; while {\em Gauge Theory of the Third Kind} is our ADG-theoresis, which is `{\em field monic, solipsistic and autonomous}'. Read on.} in the sense that the sole dynamical variable in the theory is an entirely homologically-algebraically defined $\struc$-connection acting on (the sections of) a vector sheaf $\modl$, without any recourse to or dependence on an external background geometrical differential (:$\smooth$-smooth) spacetime manifold \cite{malrap3, rap13, rap14, mall4, malzaf1}. The ADG-theoretic {\em Gauge Theory of the Third Kind}, which regards the $\struc$ connection field $(\modl ,\conn)$ as the sole dynamical variable, has been coined {\em Half-Order Formalism}\cite{malrap3, rap13, rap14}.\footnote{To distinguish it from the original {\em Second Order Formalism} of Einstein, involving the smooth spacetime metric $g_{\mu\nu}$ on a background differential spacetime manifold \cite{mtw}, as well as from the more recent  {\em First Order Formalism} of Ashtekar {\it et al.}, which, apart from a Lorentzian spin-connection $\aconn_{\mu}$, it includes the metric-{\em vierbein} $e^{a}_{\mu}$ as joint gravitational dynamical variables in the theory \cite{ash, ash1}.} {\em In ADG-GT, dynamics concerns and derives solely from the stalks (of the sheaves involved), not from the base topological space $X$ itself, which is only used for the sheaf-theoretic localisation (and continuous variation) of the `geometrical objects' (:the algebraic connection fields) that live on that surrogate external base space}.\footnote{To use a Wittgensteinian metaphor here: ``{\em Once one climbs up the ladder, one throws the ladder away}'' \cite{witt}. Similarly in our ADG-theoresis of gauge theory and gravity, {\em once we have employed the base topological space $X$ as a surrogate scaffolding for the sheaf-theoretic localisation, gauging and continuous dynamical variation of the (Einstein and Yang-Mills) gauge connection fields on it, we formulate the gauge invariant dynamics homologically-algebraically and functorially as equations involving sheaf morphisms as we showed earlier, and we then completely forget about $X$, which `atrophises' and `dissolves' in the background} (pun intended).} {\em The ADG-field dynamics is purely algebraic, smooth geometrical base spacetime manifoldless, connection field-solipsistic and autonomous}  (:dynamical connection field `self-governing' and `self-propagating').\footnote{In this respect, it has been noted and emphasised elsewhere \cite{malrap3, malrap4, rap7, rap13, rap14} 
 that the `unitary' ADG-theoretic field-pairs $\mathcal{F}=(\modl ,\conn)$ recall and conceptually resemble {\em Leibniz's Monads} \cite{leibniz}, in the sense that they are {\em autonomous (:self-governing), dynamically self-propagating and self-sustaining entities in no need of an external geometrical spacetime continuum for their dynamical support and substenance}. Moreover, in close analogy to {\em Leibniz's purely algebraic (:relational) conception of the notion of derivative in Differential Calculus} \cite{leibniz1, leibniz2}, {\em the ADG-theoretic connection fields are the sources of differentiation, hence the `causes' of dynamical variability in Mallios's theory}. Read below for more detailed discussion of {\em the close resemblance between Mallios's homological-algebraic conception of Differential Geometry in the guise of ADG and Leibniz's relational conception of derivative and, in extenso, of Differential Calculus}.}

\end{quotation}

\noindent We close this subsection by recalling from Brian Hatfield's prologue to Feynman's Lectures of Gravitation \cite{feyn} where he discusses Feynman's prophetic intuition, {\it vis-\`a-vis} Quantum Gravity, that, in a strong sense, it was quite accidental that gravity was originally formulated in terms of a metric tensor---so that Quantum Gravity would have to involve some kind of `quantising spacetime geometry'---but rather that {\em gravity fundamentally reflects some kind of deep gauge invariance}. In other words, that {\em gravity should be regarded as a gauge theory like the other three fundamental forces of Nature}:

\bigskip \noindent\hskip 0.9in
\begin{minipage}{11cm}
\noindent ``{\small ...Thus it is no surprise that Feynman would
recreate general relativity from a non-geometrical viewpoint. The
practical side of this approach is that one does not have to learn
some `fancy-schmanzy' (as he liked to call it) differential
geometry in order to study gravitational physics. (Instead, one
would just have to learn some quantum field theory.) However, when
the ultimate goal is to quantize gravity, Feynman felt that the
geometrical interpretation just stood in the way. From the field
theoretic viewpoint, one could avoid actually defining---up
front---the physical meaning of quantum geometry, fluctuating
topology, space-time foam, {\it etc.}, and instead look for the
geometrical meaning after quantization...Feynman certainly felt
that the geometrical interpretation is marvellous, `{\em but the
fact that a massless spin-$2$ field can be interpreted as a metric
was simply a coincidence that might be understood as representing
some kind of gauge invariance'}\footnote{Our emphasis of Feynman's
words as quoted by Hatfield.}...}''
\end{minipage}

\vskip 0.1in

\subsubsection{The Functoriality of 3rd Quantization in ADG-GT}

In ADG, the functoriality that pervades both the kinematical structures and the the purely gauge theoretic dynamical Einstein and Yang-Mills field equations as we expounded above, also extends naturally to include {\em geometric (pre)quantization} and {\em second (field) quantization}, processes which have also been developed entirely homologically-algebraically by ADG-theoretic means \cite{mall5, mall6, mall4, mall10, mall13, mall15, rap13, rap14}. The upshot of that {\em homological-algebraic quantisation `procedure'} is that:

\begin{quotation} 
\noindent {\bf \underline{Aphorism 3:} The ADG-Theoretic Functorial Quantum Field-Particle Duality.} The ADG-theoretic field pair $(\modl ,\conn)$ also embodies the fundamental `{\em Quantum Field-Particle Duality}' in the sense that from a geometric (pre)quantization and second quantisation vantage, {\em the (local) sections of the vector sheaves $\modl$ embodied in the ADG-theoretic connection fields represent quantum states of bosonic or fermionic field-quanta}, as follows:\footnote{The correspondences below are borrowed almost {\it verbatim} from \cite{mall5, mall6, mall4, rap13, rap14}.}

\end{quotation}

\begin{equation}\label{eq32}
\begin{array}{c}
(\modl ,\conn )\Longleftrightarrow (\mathrm{`dynamics'}, \mathrm{`kinematics'})\cr
(\conn ,\Gamma(U,\modl))\Longleftrightarrow (\mathrm{field}, \mathrm{local\, particle\, states})\cr
(\conn ,\Gamma(U, \mathcal{L}_{n=1})\Longleftrightarrow (\mathrm{Boson\, field}, \mathrm{local\, bosonic\, particle\, states})\cr
(\conn ,\Gamma(U, \modl_{n>1})\Longleftrightarrow (\mathrm{Fermion\, field}, \mathrm{local\, fermionic\, particle\, states})\cr
\end{array}
\end{equation}

\noindent Moreover, {\em the correspondences above have been seen to be purely functorial} \cite{mall5, mall6, mall4, rap13, rap14}.\footnote{Additionally, the reader should note that the fermionic sheaves $\modl_{n>1}$ in the last line above may be conventionally regarded as `{\em Odd-Grassmannian}' Sheaves \cite{freund, gosch}.}

\subsection{Demistifying and Circumventing Singularities and Field-Theoretic Unphysical Infinities.}

The $\struc$-functorial and $\struc$-invariant bottom-up {\em aufbau} of ADG has been used to totally circumvent, `deconstruct' and `demistify' singularities and other associated (non-renormalisable) unphysical infinities that have hitherto seemed to mar and assail the CDG-based Einsteinian gravity and the quantum gauge field theories of matter both of which are based on an underlying smooth geometrical spacetime manifold.

The usual physicists' consensus is that the singularities of General Relativity (GR) and the unphysical infinities of the spacetime continuum based quantum field theories of matter are indications that: 

\begin{quotation}
\noindent {\em The Laws of Physics break down, thus Nature becomes nonsensical and unpredictable, at those sites} \cite{hawk1, hawk2, hawk3, hawk4, geroch1, geroch2, wheeler1}.
\end{quotation}

With Professor Mallios we had time and again scrutinised the statement above and {\em invariably it seemed fundamentally incomprehensible to him}:

\begin{quotation}
\noindent How come the Field Laws of Physics, which we normally model after differential equations, break down at `geometrical sites' we call singularities?
\end{quotation}

It was Mallios's original and fundamental idea that:

\begin{quotation}
\noindent If we could somehow abstract, genaralise and detach the `{\em innate differential geometric mechanism}' of Calculus from its apparent inextricable dependence on a fixed $\smooth$-smooth background geometrical (spacetime) manifold, we could still do most (if not all!) of Differential Geometry without getting stuck on or breaking down at singularities or other unphysical continuous field infinities. Moreover, we could even integrate, encompass or even `absorb' singularities into the structure sheaf $\struc$ of ADG and still the whole differential geometric mechanism would hold in their very presence. That is, the $\struc$-connection fields would still define and obey differential equations (laws) in the very presence of singularities, no matter how numerous, robust or pathological those singularities might be, especially when viewed from the vantage of the Classical Differential Calculus (CDG), which is based on a background smooth geometrical spacetime manifold  \cite{mall3, mall7, mall9, mall11, malrap4, rap5, malros1, malros2}.
\end{quotation}

In other words,\footnote{And this was by far Tasos's favourite {\it motto}.}

\begin{quotation}
\noindent {\em Don't `blame' Nature for the singularities and the related unphysical infinities of the CDG-based field theories we have. Blame our Mathematics: pit it on CDG itself!}\footnote{Here, Tasos liked to use the following analogy: in much the same way as the real number line $\R$ comes short or `breaks down' when we try to solve the algebraic equation $x^{2}+1=0$, the CDG-formulated differential equations modelling the Laws of Physics appear to break down at singularities and the latter (misleadingly) appear to be {\em shortcomings and blemishes} of the Physical Laws (:differential equations) themselves. However, all we had to do is to extend $\R$ to $\com$, and the `seemingly problematic' $x^{2}+1=0$ is solved(!) {\it Mutatis mutandis} then for the extension, abstraction and generalisation of CDG to ADG: {\em from an ADG-theoretic perspective, the Field Laws of Physics do not break down at singularities} (see displayed statement below).}
\end{quotation}

\noindent Thus, by showing ADG-theoretic means that {\em the differential geometric mechanism of Calculus is inherently or innately homological algebraic} (:sheaf and category-theoretic)\footnote{More in line with Leibniz's (rather than Newton's) conception of the basic derivative operator ({\it viz}. connection) of Differential Calculus, Mallios preferred to call ADG {\em a relational theory} \cite{leibniz1, leibniz2}, where the  differential geometric mechanism derives from the algebraically modelled (dynamical) relations between the dynamical fields (connections) themselves, not from an external background geometrical spacetime continuum. Read on.} and not at all dependent on an external (to the fields) background geometrical spacetime continuum, Mallios abstracts and generalises the usual CDG to ADG and manifestly shows that:

\begin{quotation}
\noindent {\em We can actual do/use Differential Calculus in the very presence of singularities and other `CDG/differential manifold based anomalies'}, thus, {\it a fortiori}, {\em the Laws of Physics (:the differential equations between the dynamical connection fields) do not `break down' in any sense at their presence}. They still hold intact, and we can still calculate and `predict' things based on them!
\end{quotation}

We close this subsection by borrowing three quotes from \cite{malrap4}---all three all-time favourites of Tasos---that in a sense foreshadow the development of ADG and its physical applications to gravity and gauge theories of matter and, in view of our arguments in this paper, they `post-anticipate' how our purely algebraic (:ADG-theoretic) finitary, causal and quantal theoresis of vacuum Einstein Gravity and free Yang-Mills theory has come to `vindicate' them:

\noindent $\bullet$ The first two quotes by {\em Albert Einstein} are taken from the very last Appendix D of {\em The Meaning of Relativity} \cite{einst3}:

\bigskip \noindent \hskip 0.9in
\begin{minipage}{11cm}
\noindent ``{\small ...One can give good reasons why reality
cannot at all be represented by a continuous field. {\small\em
From the quantum phenomena it appears to follow with certainty
that a finite system of finite energy can be completely described
by a finite set of numbers}.\footnote{Our emphasis.} This does not
seem to be in accordance with a continuum theory, and must lead to
an attempt to {\small\em find a purely algebraic theory for the
description of reality}\footnote{Our emphasis.}...}

\hskip 0.1in

\noindent {\rm and:}

\hskip 0.1in

 {\small ...Is it conceivable that a field
theory permits one to understand the atomistic and quantum
structure of reality? Almost everybody will answer this question
with `no'. But I believe that at the present time nobody knows
anything reliable about it. {\em This is so because we cannot
judge in what manner and how strongly the exclusion of
singularities reduces the manifold of solutions. We do not possess
any method at all to derive systematically solutions that are free
of singularities}\footnote{Our emphasis.}...}''

\end{minipage}

\hskip 0.1in

\noindent $\bullet$ The third quote by {\em David Finkelstein} is taken from the introduction of his {\em Theory of Vacuum} \cite{df}:

\noindent\hskip 0.1in
\begin{quotation}
\noindent ``{\small ...The locality principle seems to catch
something fundamental about nature... Having learned that the
world need not be Euclidean in the large, the next tenable
position is that it must at least be Euclidean in the small, a
manifold.\footnote{Recall that, by definition, {\em a manifold is a locally Euclidean space}---that is to say, a space that is locally isomorphic to $\R^{n}$.}  
The idea of infinitesimal locality presupposes that the
world is a manifold.\footnote{Here, {\em the notion of infinitesimal locality mandates that the Laws of Physics be modelled after differential equations}.} 
{\small\em But the infinities of the manifold
(the number of events per unit volume, for example) give rise to
the terrible infinities of classical field theory and to the
weaker but still pestilential ones of quantum field
theory}.\footnote{Our emphasis.} The manifold postulate freezes
local topological degrees of freedom which are numerous enough to
account for all the degrees of freedom we actually observe.

The next bridgehead is {\small\em a dynamical topology, in which 
even the local topological structure is not constant but
variable}.\footnote{Our emphasis.} The problem of enumerating all
topologies of infinitely many points is so absurdly unmanageable
and unphysical that {\small\em dynamical topology virtually forces us to a
more atomistic conception of causality and space-time than the
continuous manifold}\footnote{Again, our emphasis.}...}'' 

\end{quotation}

\subsection{ADG-Field Realism, Solipsism and Monism, and Mallios's Novel Conception of Bohr's Correspondence Principle: ADG-Field `Unitarity' from 3rd Quantisation}

In the previous section, we witnessed and we argued how the close interplay between fuctoriality and $\struc$-invariance of the ADG-field dynamics is tantamount to gauge invariance. That is to say, our attempts to localise and coordinatise (:`measure') the Einstein or the Yang-Mills gauge field by employing the algebra structure sheaf $\struc$ of generalised coordinates---or `{\em arithmetics}', as Mallios preferred to call it---relative to, and localised over, a system of local gauges (:covers) $\gauge_{i}$ (of the in principle arbitrary base topological space $X$), does not affect the dynamical law that (the curvature of) the field---{\em the $\mathcal{G}\mathcal{M}$-geometric morph of the field}---obeys.\footnote{That is to say, the action functional density and the law---the differential equation derived from it by `variation' with respect to the local gauge connection potentials $\aconn$---remains invariant under $\struc$-coordinate changes and $\aut_{\struc}\modl |_{(U\in\gauge)}$ local gauge transformations.}

In turn, this means that that the field `sees through' and remains undisturbed by our `perturbing' generalised acts of measurement (:gauge localisations), hence the issue arises of what would Bohr's Correspondence Principle of the usual Quantum Theory be in our ADG-GT. Mallios had quite a fascinating, unconventional conception of, and unorthodox ideas about, that, as follows:

\begin{quotation}
\noindent Traditionally, from the very advent of Quantum Mechanics, Bohr's Correspondence Principle can be expressed as follows: observable quantum actions are represented by noncommutative `numbers' (:so-called {\em q-numbers}\footnote{Think, for instance, of Heisenberg's matrices.}), while our measurements thereof should correspond to (:should yield) commutative numbers (:so-called {\em c-numbers}---presumably, these are real numbers in $\R$, which are embedded in the complex numbers $\com$ that the usual Quantum Theory employs\footnote{In that sense, {\em real numbers are `real'} (pun intended), but over the years, the use of the real number continuum has been questioned and challenged in both Quantum Theory and Quantum Gravity (see for example \cite{ish3} for a thoughtful exposition).}).
\end{quotation}

\noindent Now, Mallios contended that {\em Bohr's c/q, commutative/noncommutative, classical/quantum dichotomy}--the {\em `classical/quantum divide'}, so to speak---is already embodied and structurally encoded in the ADG-conception of a field as the pair $(\modl ,\conn)$, in the following sense:

\begin{quotation}
\noindent Our generalised local measurements (:local `arithmetics' relative to a local gauge $U\in\gauge_{i}$) of the ADG-fields are represented by the local sections of the {\em abelian algebra} structure sheaf $\struc$ ({\it i.e.}, $\struc_{U}=\struc(U)=\Gamma(U,\struc)$).\footnote{The reader should note here a {\em key difference} between the usual conception of $c$-numbers (:results of measurements) in conventional Quantum Theory and the generalised $c$-numbers of ADG-GT. The former are sections of the constant sheaf $\cons\equiv\R\subset\com$, while the latter are sections of the `dynamically variable' sheaf $\struc$, which in turn includes anyway the constant sheaves of complex numbers ($\com$) and real  ($\R$) numbers as proper subsheaves. Mallios and Zafiris in \cite{malzaf1} give a very novel operational interpretation and physical explanation of this generalisation!} On the other hand, the noncommutativity---the $q$-number Heisenberg type of indeterminacy, the quantum fuzziness and the `quantum foam' aspect of the ADG-fields, so to speak---is already encoded in the principal group sheaf $\aut_{\struc}\modl$ of local (gauge) automorphisms of the field, which is locally isomorphic to: $\aut_{\struc}\modl (U)\simeq M^{\bullet}_{n}(\struc(U))$, the structure group sheaf of invertible $(n\times n)$-matrices, having for entries local sections of $\struc$ in $\struc |_{U\in\gauge}=\struc(U)\equiv\Gamma(U,\struc)$. Thus, $M^{\bullet}_{n}(\struc(U)$ is the ADG-field theoretic version of the `{\em local Heisenberg group}' of the theory, which is manifestly non-abelian---elements (:local sections) of which correspond to the ADG-version of $q$-numbers.
\end{quotation}

Below, we are going to give briefly a `heuristic analogy' of Mallios's seemingly unorthodox and unconventional intuition above, which has been previously noted in \cite{malrap3, rap13, rap14, malrap4}, while in the next subsection we are going to tie our heuristics below to the {\em canonical, sheaf cohomological 3rd quantisation of ADG-GT first proposed in} \cite{rap13}. 

Similar to how Bohr's Correspondence is almost tautosemous with the formal `inverse' of the original  {\em First Quantization Correspondence}:

\begin{equation}\label{eq33}
\begin{array}{c}
\mathrm{Classical\, Position\, c-number:}\; x\longrightarrow \hat{x}\,\mathrm{:Quantum\, Position\, q-number}\cr
\cr
\cr
\mathrm{Classical\, Momentum\, c-number:}\; p\longrightarrow \hat{p}\,\mathrm{:Quantum\, Momentum\, q-number}
\end{array}
\end{equation}

\noindent also by imposing the following {\em Heisenberg Uncertainty commutation relations} between the $\hat{x}$ and $\hat{p}$ operators (:matrices), which in turn generate the usual {\em Heisenberg Algebra} of traditional Quantum (Matrix) Mechanics:

\begin{equation}\label{eq34}
[\hat{x},\hat{p}]=-\imath\hbar
\end{equation}

\noindent the analogous `{\em position-momentum correspondence}' within the ADG-fields is, following \cite{malrap3, rap13, rap14, malrap4}:

\begin{equation}\label{eq35}
\begin{array}{c}
\mathrm{Local\, Particle\, `Position' \, States}\longrightarrow\mathrm{Local\, Sections\, of}\, \modl\cr
\cr
\mathrm{Local\, Field\, `Momentum'\, Operator}\longrightarrow \conn =d+\aconn
\end{array}
\end{equation}

\noindent {\em the rationale for the heuristic semantic correspondences above being that}:

\begin{quotation}
\noindent {\em Much in the same way that in conventional particle (Newtonian) mechanics velocity (speed) or momentum is (or measures) the (rate of) change of the position (state) of a particle,\footnote{Which momentum is, in turn, the differential (:derivative) of that position determination: $p:\, x\longrightarrow dx$.} in ADG-field theory, the $\struc$-connection $\conn$ in the ADG-field $(\modl ,\conn)$ is a generalised differential, acting as a sheaf morphism on the `local particle states' of the ADG-field (which are in turn represented by the local sections of $\modl$ within the ADG-field) as it were \underline{\em to change them}.}
\end{quotation}

\noindent Then, the formal `quantum deformation' (`Heisenberg uncertainty relation') of the usual Poisson Brackets Algebra of Classical Mechanics to the Heisenberg Algebra of Quantum Mechanics by the imposition of the following canonical commutation relations:

\begin{equation}\label{eq36}
\{ x,p\}=0\, \longrightarrow\, [\hat{x},\hat{p}]=-\imath\hbar
\end{equation}

\noindent can heuristically be cast within the ADG-field as follows:

\begin{equation}\label{eq37}
[\modl ,\conn]=\aut_{\struc}\modl\, \stackrel{\mathrm{locally}}{\longrightarrow}\, [s\in\modl(U), d+\aconn]=M_{n}(\struc_{U})s=M_{n}(\struc(U))s
\end{equation}

\noindent where $s\in\modl_{\struc}(U)$ (with $\modl_{\struc}(U)\equiv\Gamma(U,\modl_{\struc})\simeq\struc^{n}(U)$) a `local quantum particle state' within the ADG-field $(\modl ,\conn)$ and $M_{n}(\struc(U))$ a local section of the structure group sheaf $\mathcal{G}=\aut_{\struc}\modl$ of the ADG-field relative to the local open gauge $U$.

The `canonical'  commutator `auto-uncertainty' relation within the ADG-field $(\modl ,\conn)$ in equation (\ref{eq37}) above can then be heuristically interpreted as follows:

\begin{quotation}
\noindent The `canonical' commutator quantum uncertain relation between the generalised local position quantum particle states in $\modl$ and its dua; generalised momentum field operator (:sheaf morphism) $\conn$ generates and induces a dynamical local gauge transformation in $\aut_{\struc}\modl |_{U}=M^{n}(\struc(U))$, which then acts on the local quantum particle states to change them and, as it were, to `blur' them (:`quantum fuzziness' or `quantum foam').
\hskip 0.1in
\noindent In this heuristic sense, $\modl$ is `complementary' to $\conn$, thus the ADG-field may be thought of as being `{\em self-complementary}' and `{\em self-quantum}'. In this sense we argued in \cite{malrap3, rap13, rap14, malrap4} that {\em the ADG-field is an already self-quantum, auto-dynamical entity}.
\end{quotation}

We can formalise the generalised ADG-theoretic quantum uncertainty/complementarity relation above as a homological $\struc$-tensor product morphism type of map:

\begin{equation}\label{eq38}
\mathcal{Q}:\,\modl_{\struc}\otimes_{\struc}\conn\longrightarrow\aut_{\struc}\modl_{\struc}
\end{equation}

\noindent thus finally arrive at the definition of the {\em `unitary' quantal ADG-gauge field}\footnote{The reader should note that, issuing from the 3rd Quantisation scheme for ADG-field theory originally presented in \cite{rap13}, the epithets {\em unitary quantum} carry standard meaning in the usual Quantum Theory, hence we use inverted single quotes around the word {\em unitary}, while instead of {\em quantum} we use {\em quantal} throughout our work, in order to avoid confusion of semantic reference.} as being the following tetrad:

\begin{equation}\label{eq39}
\mathbf{U}:=(\modl ,\conn, \aut_{\struc}\modl , \mathcal{Q})
\end{equation}

\noindent which encodes the following four pieces of important information:

\begin{enumerate}

\item The vector sheaf $\modl$ of generalised quantum particle `position' states;

\item The connection field $\conn$, which effectuates dynamical `momentum' field-like changes of the said states by acting as a sheaf morphism on $\modl$'s local sections;

\item The structure gauge group sheaf $\aut_{\struc}\modl$ of local gauge transformations, also effectively acting on $\modl$ as its principal structure group sheaf\footnote{And conversely, $\modl$ can be viewed as the associated sheaf to the principal group sheaf $\aut_{\struc}\modl$.}; and finally,

\item The quantum uncertainty operator (:morphism) $\mathcal{Q}$, which stands for an $\aut_{\struc}\modl$-valued quantum act of perturbation on the dynamical action of the connection field on the local quantum particle states (:local sections) of $\modl$.

\end{enumerate}

\noindent The epithet `{\em unitary}' for $\mathbf{U}$ in (\ref{eq39}) above indicates that it is a holistic entity, an inseparable whole, encoding state ($\modl$), dynamical changes of state ($\conn$), gauge symmetries and invariances of dynamical changes of state ($\aut_{\struc}\modl$), and quantum uncertainty of `determination' ($\mathcal{Q}$), {\em all-4-in-1}. 

\begin{quotation}
\noindent {\em All there is in ADG-field theory are the Leibnizian Monad type of entities} $\mathbf{U}=(\modl ,\conn, \aut_{\struc}\modl , \mathcal{Q})$, {\em nothing else}. This is what was referred to in \cite{malrap3, rap13, rap14, malrap4} as `{\em ADG-field solipsism and monism}'.
\end{quotation}

\noindent Thus, perhaps more importantly, the adjective `unitary' given to the tetrad $\mathbf{U}=(\modl ,\conn, \aut_{\struc}\modl , \mathcal{Q})$ above pertains to the fact that in ADG-field theory, {\em there is no reference and recourse to, no dependence whatsoever on, an external (to the ADG-fields themselves) background geometrical spacetime manifold}. It follows that the usual distinction and schism {\em internal/external} that is normally reserved for the {\em symmetries} of the usual gauge field theories of matter---whether classical or quantum---loses its meaning in our ADG-theoresis of Vacuum Einstein Gravity and Free Yang-Mills theories.\footnote{Traditionally, we reserve the epithet `external' for `the external (to the fields) spacetime symmetries', while `internal' is normally reserved for gauge degrees of freedom and their symmetries \cite{au, gosch}.} This has also been succinctly pointed out more recently in \cite{malzaf1}.

\subsubsection{Interregnum: Drawing Formal Links with 3rd Quantisation}

In the discussion below, we briefly draw links between the formal `canonical' quantisation heuristics above and the functorial sheaf cohomological 3rd Quantisation of ADG-GT scenario presented in \cite{rap13}.

To that end, we recall that in \cite{rap13} we intuited that since the ADG-fields $\mathcal{F}=(\modl ,\conn)$ are dynamically
self-supporting, autonomous monadic entities as we emphasised earlier, and 
moreover, since they are `{\em self-dual}'\footnote{In the sense that the connection momentum-like field $\conn$ is quantum dual to the `position quantum particle states' (represented by the local sections of) $\modl$.}

\begin{quotation}
\noindent {\em a possible quantization scenario for them should
involve solely their two constitutive parts, namely, $\modl$ and
$\conn$, without recourse to/dependence on extraneous structures
({\it e.g.}, a base spacetime manifold) for its mathematical support and its self-consistent (physical)
interpretation}.
\end{quotation}

Thus, in what formally looked like a canonical quantization-type of
scenario,

\begin{quotation}
\noindent {\em in} \cite{rap13} {\em we envisaged abstract non-trivial local commutation
relations between the abstract position} (:$\modl$) {\em and
momentum} (:$\conn$) {\em aspects of the ADG-fields}.
\end{quotation}

To that end, we recalled that

\begin{quotation}
\noindent {\em there are certain local (:differential) forms that
uniquely characterize sheaf cohomologically the vector sheaf
$\modl$ and the connection $\conn$ parts of the ADG-fields}
$\mathcal{F}=(\modl ,\conn)$
\end{quotation}

Thus, the basic heuristic-intuitive idea in \cite{rap13} was to identify the relevant
forms and then posit non-trivial commutation relations between
them. Moreover, for the sake of the aforementioned `{\em dynamical
ADG-field autonomy}', we would like to require that

\begin{quotation}
\noindent the envisaged commutation relations should not only
involve just the two components ({\it i.e.}, $\modl$ and $\conn$) of
the total ADG-fields $\mathcal{F}=(\modl ,\conn)$, but they should also somehow  {\em
`algebraically close' within the fields themselves}---{\it i.e.}, the result of
their commutation relations should not take us `outside' the total
ADG-field structure (and its `dynamical auto-transmutations'), which anyway
is the only dynamical structure involved in our
theory.\footnote{This loosely reminds one of the theoretical
requirement for algebraic closure of the algebra of quantum
observables in canonical QG, with the important difference however
that the $\mathrm{Diff}(M)$ group of the external (to the
gravitational field) spacetime manifold must also be considered in
the constraints, something that makes the desired closure of the
observables' algebra quite a hard problem to overcome
\cite{thiem2}. In \cite{malrap3, rap13, rap14} we discuss certain
difficult problems that $\mathrm{Diff}(M)$ creates in various QG
approaches, as well as how its manifest absence in ADG-gravity can
help us bypass them totally. For, recall that from the
ADG-perspective gravity is an external (:background) spacetime
manifold unconstrained (because it is a background spacetime manifoldless) pure gauge theory (:of the 3rd kind).}
\end{quotation}

Keeping the theoretical requirements above in mind, we recall from
\cite{mall1, mall2, mall5, malrap2, mall4} {\em two important sheaf
cohomological results}:

\begin{enumerate}

\item That, sheaf cohomologically, the vector sheaves $\modl$ are
completely characterized by a so-called {\em coordinate
$1$-cocycle} $\phi_{\alpha\beta}\in Z^{1}(\gauge
,{\mathcal{G}}{\mathcal{L}}(n,\struc))$ associated with any system
$\gauge$ of local gauges of $\modl$. Intuitively, this can be
interpreted in the following `Kleinian symmetry-geometry' way: since any (vector)
sheaf is completely determined by its (local) sections,\footnote{A
basic motto (:fact) in sheaf theory is that ``{\em a sheaf is its
sections}'' \cite{bredon, mall1}. If we know the local data (:sections),
we can produce the whole sheaf space by restricting and collating
them relative to an open cover $\gauge$ of the base topological
space $X$. This is the very process of `sheafification' (of a
preasheaf) \cite{bredon, mall1}.} one way of knowing the latter is to know
how they transform---in passing, for example, from one local gauge
($U_{\alpha}\in\gauge$) to another ($U_{\beta}\in\gauge$), with
$U_{\alpha}\cap U_{\beta}\neq\emptyset$ and $\gauge$ a chosen
system of local open gauges covering $X$.\footnote{In particular,
$\phi_{\alpha\beta}$ can be locally expressed as the $\struc
|_{U_{\alpha\beta}}$-isomorphism:
$\phi_{\alpha}\circ\phi_{\beta}^{-1}\in\aut_{\struc_{\alpha\beta}}
(\struc^{n}|_{U_{\alpha\beta}})=\mathrm{GL}(n,\struc(U_{\alpha\beta}))={\mathcal{G}}{\mathcal{L}}(n,\struc)(U_{\alpha\beta})$,
in which expression the familiar {\em local coordinate transition
(:structure) functions} appear. Hence, also the `natural' structure
(:gauge) group sheaf
$\aut_{\struc}\modl={\mathcal{G}}{\mathcal{L}}(n,\struc)$ of $\modl$
arises.} To know something ({\it e.g.}, a `space') is to know how
it transforms, the fundamental idea underlying Klein's general
conception of `geometry' \cite{klein}. 

Thus, the bottom-line here is that the
characteristic cohomology classes of vector sheaves $\modl$ are
completely determined by $\phi_{\alpha\beta}$; write:

\begin{equation}\label{eq40}
[\phi_{\alpha\beta}]\in
H^{1}(X,{\mathcal{G}}{\mathcal{L}}(n,\struc))=\lim_{\overrightarrow{\;\;\gauge\;\;}}H^{1}(\gauge,
{\mathcal{G}}{\mathcal{L}}(n,\struc))
\end{equation}

\noindent where the $\gauge$s, normally assumed to be {\em locally
finite open coverings of} $X$
\cite{mall1, mall2, malrap1, malrap2, malrap3}, constitute a {\em
cofinal subset} of the set of all proper open covers of
$X$.\footnote{An assumption that has proven to be very
fruitful in applying ADG to the formulation of a locally finite,
causal and quantal Vacuum Einstein Gravity and Free Yang-Mills theories, 
as we argued in the first part of this paper and throughout 
our past works \cite{malrap1, malrap2, malrap3, rap5, rap7, rap11, rap13, rap14}. {\it En passant}, we also note that in \cite{rap13}
the direct (:inductive) limit depicted in (\ref{eq40}) above
is secured by the `cofinality' of the set of finitary (:locally
finite) open coverings of $X$ that we choose to employ
\cite{sork0, rapzap1, rapzap2, rap1, rap2, malrap1, malrap2, malrap3, rap5, rap7} and it was emloyed $K$-theoretically to link the 3rd Quantisation of ADG-fields scenario with Mallios's $K$-theoretic musings on topological algebra structure sheaves $\struc$ and the 2nd Quantisation classification of the local quantum particle states (:local sections) of the vector sheaves involved in the ADG-fields into Bosons (:$\modl$ is a line sheaf $\mathcal{L}_{n=1}$) and Fermions (:$\modl$ is a vector sheaf of rank $n>1$) \cite{mall6}.}
{\it In toto}, we assume that $\phi_{\alpha\beta}$ {\em encodes all the
(local) information we need to determine the local
quantum-particle states of the field in focus} ({\it i.e.}, the local
sections of $\modl$).

\item On the other hand, it was observed in \cite{rap13} that {\em locally
$\conn$ is uniquely determined by the so-called `gauge potential'
$\aconn$}, which is normally ({\it i.e.}, in CDG) defined as a Lie
algebra (:vector) valued $1$-form \cite{gosch}. Correspondingly, in ADG
$\aconn$ is seen to be an element of
$M_{n}(\Omega(U))=M_{n}(\Omega)(U)=\Omega(\modl
nd\modl)$,\footnote{Note that, as also mentioned earlier, in ADG by definition one has: $\Omega :=\modl^{*}:=\mathcal{H}om_{\struc}(\modl  ,\struc)$. That is,
the $\struc$-module sheaf $\Omega$ of abstract differential
$1$-forms is dual to the vector sheaf $\modl$, much like in the
classical theory (:CDG of $\smooth$-smooth manifolds) where differential
forms (:cotangent vectors) are dual to tangent vectors \cite{gosch}, although
as it has been empasised throughout our works, in ADG the epithet
`(co)tangent' is meaningless due to the manifest absence of an
operative background space(time) of any kind (and especially of a
base manifold).} thus it is called {\em the local
$\struc$-connection matrix $(\aconn_{ij})$ of $\conn$, with
entries local sections of $\modl^{*}=\Omega$}. In turn, this means
that locally $\conn$ splits in the familiar way, as follows:

\begin{equation}\label{eq41}
\conn = d +\aconn
\end{equation}

\noindent where $\partial$ is the usual `inertial' (:flat)
differential\footnote{As noted in the previous section, in ADG, the Cartan-K\"ahler differential $d$, like $\conn$, is defined
as a linear, Leibnizian $\cons$-sheaf morphism $d :\;
\struc\rightarrow\omg$, thus it is an instance of on
$\struc$-connection; albeit, a  {\em flat} one
(:$\curv(d)=d^{2}=0$), which is secured by the very definition of curvature in (\ref{eq17}) and the niloptency of the exterior differential $d$.} and $\aconn$ the said
gauge potential. In ADG-gravity, the \texttt{total field} $\conn$
as a whole (:`globally') represents the {\em gravito-inertial
field}, but locally it can be separated into its inertial
(:$d$) and gravitational (:$\aconn$) parts.\footnote{In the classical theory of gravity (:General Relativity; abbr. GR), the physical meaning of this local separation
of the \texttt{total field} $\conn$ into $\partial$ and $\aconn$ reflects the {\em local principle of equivalence}; namely, that {\em locally, the spacetime manifold $M$ of GR is flat Minkowski space}, or equivalently, that {\em locally, GR reduces to Special Relativity}, or perhaps more importantly, that {\em gravity can always be `gauged away' locally by a suitable choice of `gauge'} (:local inertial frame). This is simply Einstein's elevator gedanken experiment \cite{mtw}.}

\end{enumerate}

Thence, the envisaged sheaf cohomological canonical
quantization-type of scenario for the total ADG-fields
$\mathcal{F}=(\modl ,\conn)$ rests essentially on positing the
following non-trivial abstract Heisenberg-type local commutation
relations between (the characteristic forms that completely
characterize) $\modl$ (:abstract `{\em position}' particle states) and
$\conn$ (:abstract `{\em momentum}' field operator). Thus, heuristically
we posited in \cite{rap13} the following `canonical' commutation relations:

\begin{equation}\label{eq42}
[\phi ,\conn ]\stackrel{loc.}{=}[\phi_{\alpha\beta},d
+\aconn_{ij}]_{U_{\alpha\beta}}=[\phi_{\alpha\beta},d
]_{U_{\alpha\beta}}+[\phi_{\alpha\beta},\aconn_{ij}]_{U_{\alpha\beta}}
\end{equation}

\noindent stressing also that, as highlighted in \cite{rap13},

\begin{quotation}
\noindent the local commutation relations in (\ref{eq42}) above are
well defined, since they effectively {\em close within the
noncommutative $(n\times n)$-matrix Klein-Heisenberg algebra}
$\modl
nd\modl(U_{\alpha\beta})=M_{n}(\struc(U_{\alpha\beta}))=M_{n}(\struc)(U_{\alpha\beta})$
of the field's endomorphisms---the field's `noncommutative
Kleinian geometry' we mentioned earlier representing what Mallios intuited as some kind of 
`quantum field foam'---the {\em intrinsically noncommutative aspect of the ADG-fields}.
\end{quotation}

\noindent This `{\em algebraic closure}' is in accord with the
theoretical requirement we imposed  earlier, namely that,

\begin{quotation}
\noindent the abstract, Heisenberg-like, canonical quantum
commutation relations between the two components $\modl$ and
$\conn$ of the ADG-fields should not take us outside the fields,
but should rather {\em close} within them.\footnote{Here, one
could envisage an abstract Heisenberg-type of algebra freely
generated (locally) by $\phi$ (:abstract position) and $\aconn$
(:abstract momentum), modulo the (local) commutation relations
(\ref{eq41}). Plainly, it is a subalgebra of $\modl nd\modl(U)$,
but deeper structural investigations on it must await a more
complete and formal treatment.}
\end{quotation}

\noindent Indeed, $\modl nd\modl$ {\em is precisely the algebra
sheaf of internal/intrinsic (dynamical) self-transmutations of the
(quantum particle states of the) field}---by definition, the
$\modl$-endomorphisms in ${\mathcal{H}}om_{\struc}(\modl ,\modl)$
(:quantum field foam). 

This is another aspect of {\em the quantum
dynamical autonomy of ADG-fields}:

\begin{quotation}
\noindent the $\modl$ (:abstract point-particle/position) part of
the ADG-field is `{\em complementary}', in the quantum sense of
`complementarity', to $\conn$ (:abstract field-wave/momentum).
Thus, {\em the total ADG-fields $\mathcal{F}=(\modl .\conn)$ are `quantum self-dual'
entities} \cite{malrap3, malrap4, rap5, rap13}.\footnote{From our abstract and background spacetime manifoldless perspective, the de
Broglie-Schr\"odinger wave-particle duality is almost tautosemous
with the Bohr-Heisenberg momentum-position complementarity.}
\end{quotation}

Furthermore, by choosing $\phi_{ab}=\phi_{ab}^{in}$\footnote{The
superscript `$in$' stands for `{\em inertial}', and it represents
a choice (:{\em our} choice!) of a local
change-of-gauge
$\phi^{in}_{\alpha\beta}\in\gl(n,\struc)_{\alpha\beta}\equiv\Gamma(U_{\alpha\beta},\gl(n,\struc)$
that would take us to a locally inertial frame of $\modl$ over
$U_{\alpha\beta}\subset X$.} so that $\aconn$ is `gauged
away'---{\it i.e.}, by setting $\aconn=0$,\footnote{As noted earlier, this is an
analogue of the Equivalence Principle (EP) of GR in ADG-gravity,
corresponding to the local passage to an `inertial frame' (:one
`covarying' with the gravitational field; {\it e.g.}, recall
Einstein's free falling elevator {\it gedanken} experiment) in
which the curved gravito-inertial $\conn$ reduces
to its flat `inertial' $\struc$-connection part $d$
\cite{mall1,mall2,malrap1,malrap2,malrap3}. As noted above, this just reflects the
well known fact that {\em GR is locally SR}, or conversely, that
when SR is localized ({\it ie}, `gauged' over the base spacetime
manifold) it produces GR (equivalently, the curved Lorentzian
spacetime manifold of GR is locally the flat Minkowski space of
SR). {\it In summa}, gravity (:$\aconn$) has been locally gauged
away, and what we are left with is the inertial action $d$
of the ADG-gravitational field $\conn$. It must be also stressed
here that the choice of a locally inertial frame, like all gauge
choices, is an externally imposed constraint in the
theory---`externally', meaning that it is {\em we}, the external
(to the field) experimenters/theoreticians (`observers') that we
impose such constraints on the field ({\it i.e.}, we make choices
about what aspects of the field we would like to single out and,
ultimately, observe/study).} reduces (\ref{eq41}) to (omitting the
local open gauge indices/subscripts `$\alpha ,\beta$'):

\begin{equation}\label{eq43}
[\phi^{in} ,d ]=\phi^{in}\circ d
-d\circ\phi^{in}
\end{equation}

\noindent Moreover, since we are sheaf cohomologically guaranteed
that $d\circ\phi=0$ globally, which is tantamount to the
very {\em existence} of an $\struc$-connection $\conn$ (globally)
on $\modl$ \cite{mall1, mall2, mall4},\footnote{This essentially
corresponds to the fact that the coordinate $1$-cocycle
$\phi_{\alpha\beta}\in Z^{1}(\gauge ,\Omega)$ is actually a {\em
coboundary} (:a closed form), belonging to the zero cohomology
class $[d\phi_{\alpha\beta}]=0\in H^{1}(X,M_{n}(\Omega))$,
which in turn guarantees the existence of an $\struc$-connection
on $\modl$ as the so-called {\em Atiyah class} $\mathbf{A}$ of
$\modl$ vanishes
(:$\mathbf{A}(\modl):=[d\phi_{\alpha\beta}]=0$)
\cite{mall1, mall2, mall4}.} (\ref{eq42}) further reduces to:

\begin{equation}\label{eq44}
[\phi^{in},d ]=\phi^{in}\circ d
\end{equation}

\noindent Now, a heuristic physical interpretation can be given to
(\ref{eq43}) if we consider its effect (:action) on a local
section
$s\in\modl_{\alpha\beta}:=\modl(U_{\alpha\beta})\equiv\modl
|_{U_{\alpha\beta}}$:

\begin{equation}\label{eq45}
[\phi^{in} , d
](s)=(\phi^{in}\circ d)(s)=\phi^{in}(d s)
\end{equation}

\noindent (\ref{eq44}) designates the inertial dynamical action of
$\conn$ ({\it i.e.}, the action of its locally flat, inertial part
$d$) on (an arbitrary) $s$, followed by the gauge
transformation of $d s$ to an inertial frame
$e_{in}^{U_{\alpha\beta}}\subset\modl_{\alpha\beta}$ `{\em
covarying}' with the inertio-gravitational field. 

It may be interpreted as expressing
what happens to a `vacuum graviton state' $s$ when it is first
acted upon\footnote{Recall that we are considering only {\em
vacuum} gravity, in which the non-linear gravitational field
`couples' solely to itself(!)} by the inertial part of the
\texttt{total} ADG-gravitational field $\conn$ and
then\footnote{The sequential language used here should not be
interpreted in an temporal-operational sense---as it were, as
`operations carried out sequentially in time'.} to an inertial
frame that in a sense `{\em covaries}' with the said inertial
change $d$ of $s$.

Heuristically, we further intuited in \cite{rap13} that one can perhaps get a more adventurous (meta)physical insight into
(\ref{eq44}) by defining the {\em uncertainty operator} $\unc$ as

\begin{equation}\label{eq46}
\unc :=\phi^{in}\circ\partial\in\modl nd\modl
\end{equation}

\noindent and by delimiting all the quantum-particle (:abstract
position) states of the field (:local sections of $\modl$) that
are annihilated by it. Intuitively, these are formally the local
`{\em classical-inertial}' states

\begin{equation}\label{eq47}
\modl^{cl}_{U}:=\mathrm{span}_{\mathbf{\com}}\{ s\in\modl(U) :\;
\unc(s)=0\}=:\mathrm{ker}(\unc)
\end{equation}

\noindent for which the abstract sheaf cohomological Heisenberg
uncertainty relations (\ref{eq42}) vanish. Plainly,
$\modl^{cl}(U)$ is a $\cons$-linear subspace of
$\modl(U)$---{\em the kernel of} $\unc$.

On top of the above, intuitively it makes sense to assume that
$\unc$ is a `projector'---a primitive idempotent (:projection
operator) locally in $\modl nd\modl$ ({\it i.e.}, in
$M_{n}(\struc(U))$)---since the `gedanken' operation of `{\em
inertially covarying with a chosen local inertial frame}' must
arguably be idempotent.\footnote{After all, `{\em inertially
covarying the inertial state leaves it inertially covariant}'. Or,
to use a famous Einstein `{\it gedanken} metaphor': `{\em jumping
on a light-ray (in order to ride it) twice, simply leaves you
riding it}'(!)} This means that $\unc^{2}=\unc$, so that $\unc$
separates (chooses or projects out) the `classical'
($\mathrm{eigen}_{0}(\unc)\equiv\mathrm{ker}(\unc)$) from the
quantum ($\mathrm{eigen}_{1}(\unc)$) local quantum
gravito-inertial states.\footnote{In \cite{rap13}, a formal mathematical reason why we chose $\unc$ to be
a projection operator was to apply it and relate our 3rd Quantisation scenario to Mallios's $K$-theoretic perspective on 2nd (Field) Quantisation in \cite{mall6}.}.

Finally, in line with \cite{rap13}, we would like to ask {\it en passant} here the following
highly speculative question:

\begin{quotation}
\noindent Could the generation/emergence of
(inertio-gravitational) mass be somehow accounted for by a
(spontaneous)  symmetry breaking-type of mechanism, whereby, the
dynamical automorphism group $\aut_{\struc}\modl$ of the ADG-gravitational
field $(\modl ,\conn)$ reduces to its subroup that leaves
$\mathrm{ker}(\unc)$ invariant? Alternatively intuited, could the
emergence of inertio-gravitational mass be thought of as the
result of some kind of `quantum anomaly' of 3rd-quantized
Vacuum Einstein Gravity?\footnote{The epithet `quantum' adjoined to `anomaly' is
intended to distinguish the effect intuited above from the usual
anomalies. A `quantum anomaly' is the `converse' of an anomaly in
the usual sense, in that what was a symmetry of the {\em quantum}
theory (:an element of $\aut_{\struc}\modl$ in our case) ceases to be a
symmetry of the `classical domain' of our theory
(:$\mathrm{ker}(\unc)$). Let it be stressed that the emergence of
gravito-inertial `mass' in the sense intuited here has a truly
relational (:algebraic) and `global' flavour reminiscent of Mach's
ideas: `global' gravitational field symmetries in $\aut_{\struc}\modl$ are
locally reduced to inertial ones, and sheaf theory's ability to
interplay between local and global comes in handy in this respect
\cite{malzaf1}. (In \cite{malzaf1}, Mallios and Zafiris do a great job in highlighting exactly how sheaf theory
allows one to transit from `local' to `global', and vice versa.)}
\end{quotation}

\noindent $\bullet$ {\bf Identifying 3rd Quantisation within our `Unitary' Quantal ADG-Gauge Field Tetrads.} The alert reader must have already noticed that the `canonical' commutation relations (\ref{eq42}) are the sheaf cohomological versions of our `heuristic' canonical commutation relations (\ref{eq37}). Also, by comparing the commutator expressions (\ref{eq37}) and (\ref{eq43}) and the associated definitions of the operators (:morphisms) $\mathcal{Q}$ in (\ref{eq38}) and $\unc$ in (\ref{eq46}), the astute reader must have realised that $Ran(\mathcal{Q})\subset Ran(\unc)$, as $M_{n}^{\bullet}(\struc(U))\subset\modl nd\modl |_{U}=M_{n}(\struc(U))$. 

Overall, and without loss of generality of mathematical structures or physical interpretation thereof, in the light of 3rd Quantisation our `Unitary' Quantal ADG-Gauge Field Tetrads in (\ref{eq39}) can now be identified with the tetrad:

\begin{equation}\label{eq48}
\mathbf{U}:=(\modl ,\conn, \aut_{\struc}\modl , \mathcal{Q})\equiv(\modl ,\conn, \modl nd\modl , \unc)
\end{equation}

\noindent which carries the same denomination as before {\em `Unitary' Quantal ADG-Gauge Field} and is again symbolised by $\mathbf{U}$.

Now, by recalling the meaning we ascribed to $\mathbf{U}$ as {\em an inseparable (:`unitary'), self-dual, 3rd-quantised, auto-dynamical, background spacetime manifoldless gauge field of the 3rd kind}, we can, {\it mutatis mutandis} address the traditional distinction and schism in the usual quantum theory between `observer' (classical exosystem) and `observed' (quantum endosystem) \cite{wheeler, df}.\footnote{This pertains to the (in)famous {\em Heisenberg scnitt } (:Heisenberg cut): the schism that divides and separates the classical from the quantum phenomena and it delimits the boundary across which the wave function supposedly collapses upon measurement. It is when $q$-numbers become $c$-numbers, and probability amplitudes become probability distributions. See \cite{wheeler} for a plethora of classic articles on the Heisenberg Schnitt and the Quantum Theory of Measurement. It's where John Wheeler noted that ``{\em No phenomenon is a phenomenon unless it is an observed phenomenon}'' \cite{wheeler}.} We are not going to elaborate in detail on this subtle and important point here,\footnote{We are going to tackle it further in \cite{malrap4}.} but it is noteworthy to mention that significant work has been done on defining by the homological algebraic (:category and topos-theoretic) means of ADG internally consistent {\em quantum observables} in the theory, without recourse to any external spacetime manifold, as befits the `unitary' and quantal ADG-gauge field theory \cite{zaf1, zaf2, malzaf1, malzaf1}.

In the light of the physical interpretation of the $\mathbf{U}$ ADG-field tetrads above, we conclude this subsection by quoting David Finkelstein from \cite{df}, making some protphetic remarks about {\em the future of physical laws} {\it vis-\`a-vis} his Quantum Relativity Theory approach to Quantum Gravity, based on `abolishing' this external/internal field distinction and schism of the usual theory:\footnote{This quote was one of Tasos's all-time favourites, and I recall fondly him urging me to include it in the opening talk of the first {\em Glafka 2004: Iconoclastic Approaches to Quantum Gravity} international theoretical physics conference that we jointly organised in Athens, Greece---setting thus `{\em The Spirit of the Meeting}' \cite{rap11}.}

\begin{quotation}
\noindent ``...{\small\em What are we after as physicists? Once I
would have said, the laws of nature; then, the law of nature. Now
I wonder.}\footnote{Our emphasis.}

{\small A law, or to speak more comprehensively, a theory, in the
ordinary sense of the word, even a quantum theory of the kind
studied today by almost all quantum physicists, is itself not a
quantum object. We are supposed to be able to know the theory
completely, even if it is a theory about quanta. Its symbols and
rules of inference are supposed to be essentially non-quantum. For
example, ordinary quantum theory assumes that we can know the form
of the equations obeyed by by quantum variables exactly, even
though we cannot know all the variables exactly. This is
considered consistent with the indeterminacies of quantum theory,
because the theory itself is assumed to sum up conclusions from
arbitrarily many experiments.

Nevertheless, since we expect that all is quantum, we cannot
consistently expect such a theory to exist except as an
approximation to a more quantum conception of a theory. At present
we have non-quantum theories of quantum entities. Ultimately the
theory too must reveal its variable nature. For example, the
notion that an experiment can be repeated infinitely often is as
implausible as the notion that it can be done infinitely quickly
($c=\infty$), or infinitely gently ($\hbar=0$).

It is common to include in the Hamiltonian of (say) an electron a
magnetic field that is treated as a non-quantum constant,
expressing the action of electric currents in a coil that is not
part of the endosystem but the exosystem. Such fields are called
external fields. Upon closer inspection, it is understood, the
external field resolves into a host of couplings between the
original electron and those in the coil system, now part of the
endosystem.

{\em It seems likely that the entire Hamiltonian ultimately has
the same status that we already give the external field. No
element of it can resist resolution into further quantum
variables. In pre-quantum physics the ideal of a final theory is
closely connected with that of a final observer, who sees
everything and does nothing. The ideal of a final theory seems
absurd in a theory that has no final observer. When we renounce
the ideal of a theory as a non-quantum object, what remains is a
theory that is itself a quantum object. Indeed, from an
experimental point of view, the usual equations that define a
theory have no meaning by themselves, but only as
information-storing elements of a larger system of users, as much
part of the human race as our chromosomes, but responding more
quickly to the environment. The fully quantum theory lies
somewhere within the theorizing activity of the human race itself,
or the subspecies of physicists, regarded as a quantum system. If
this is indeed a quantum entity, then the goal of knowing it
completely is a Cartesian fantasy, and at a certain stage in our
development we will cease to be law-seekers and become law-makers.

It is not clear what happens to the concept of a correct theory
when we abandon the notion that it is a faithful picture of
nature. Presumably, just as the theory is an aspect of our
collective life, its truth is an aspect of the quality of our
life}\footnote{Again, our emphasis throughout.}...}''

\end{quotation}

\section{Brief Philosophical Epilegomena: Anastasios Mallios's Original Vision and Posthumous Future Legacy}

This author's earliest recollections of exchanges with Professor Anastasios Mallios in the late 1990s/early 2000s during multiple dinner evenings at a little known, quaint and cosy little tavern, fittingly called {\it Algebra}, situated in the northern Athens suburb of Paleo Psychiko, about {\em the potential import of Abstract Differential Geometry in current persistently un(re)solved technical (:mathematical), conceptual-cum-semantic and philosophical issues in Quantum Gravity and Quantum Gauge Theory research}, focused mainly on two fronts:

\begin{enumerate}

\item  The algebraic essence and origin of `physical space' and its `physical geometry' and, as a `result':

\item The non-existence of an {\it a priori} `geometrical space(time)', but quite on the contrary, the emergence of `geometrical space(time)' as an outcome of an algebraic (:relational) dynamics (:dynamical interactions) between the `physical geometrical objects' (:the physical fields) that live on a surrogate and virtual `space'.

\end{enumerate}

\noindent Focusing on the two items above, below I will try to recall and `reconstruct' the origins and motivations of ADG.

\subsection{The Original Vision: `Geometrical Space(-Time)' comes from `Algebraic Dynamics'} 

From numerous exchanges, close collaboration and warm friendship with the creator of ADG over more than one and a half decades, this author maintains and is willing to argue that Professor Mallios's inspired {\it magnum opus} (ADG) was originally motivated by two main aspects---one technical-mathematical, the other intuitive-heuristic-conceptual and physical---that synergistically feed each other, grow holistically together, and almost perfectly dovetail with each other in the {\it aufbau} of ADG..

\subsubsection{Tracing the Mathematical `Origins' of ADG: the Categorical Duality between Algebra and Geometry}

Very early on in my crossing worldlines with Professor Mallios, he kept on bringing up one of his all-time favourite quotes by Sophie Germain \cite{germain} as being a motto at the very heart of ADG: 

\begin{quotation}

\noindent ``{\em G\'eometrie est une Alg\`ebre bien figur\'e, mais Alg\`ebre est une G\'eometrie bien \'ecrite}''\footnote{English translation: ``{\em Geometry is a well figured (or designed) Algebra, while Algebra is a well written Geometry}''.}

\end{quotation}

\noindent The quote above perfectly encapsulates, in a beautifully poetic way, {\em the fundamental mathematical (:categorical) duality between Algebra and Geometry}, which, in modern mathematical (:category-theoretic) parlance may be boiled down and reduced to two cornerstone results, both of which we have played a central role in Mallios's developing his theory:\footnote{Tasos Mallios in numerous private communications.}

\begin{enumerate}

\item \underline{\bf Gel'fand Duality and the Gel'fand-Stone Theorem:} Generally and loosely speaking, {\em Gel'fand duality is a general duality between spaces and algebras of functions defined on them}. In particular, for the case of {\em compact topological spaces} and {\em abelian $C^{*}$-algebras}, Gel'fand duality roughly pertains to the result that {\em every commutative $C^{*}$-algebra $A$ is equivalent to the abelian $C^{*}$-algebra of continuous functions on a suitably and `naturally topologised' (:using the algebraic structure itself)\footnote{The set of the algebra's idecomposable, irreducible atomic elements so to speak---its (primitive or prime) ideals.} space called its Gel'fand Spectrum $Spec(A)$} \cite{gelfand, johnstone, landsman}.\footnote{Earlier in the present paper, we witnessed an instance of {\em `discrete' Gel'fand Duality}, when we discussed {\em Sorkin's finitary poset discretisations of locally compact continuous manifolds and their Gel'fand-dual incidence Rota algebras} (cf. Section 1). It must be emphasised here that the adjective `{\em equivalent}' in the statement of Gel'fand Duality above pertains to {\em the categorical equivalence (:functorial correspondence) between the category of abelian $C^{*}$-algebras and that of (compact) topological spaces}.} 

For Tasos, the main technical and conceptual essence of the mathematical result above is that `{\em Space (Geometry/Topology) can be somehow `derived' or extracted from Algebraic Structure}'.\footnote{Again, Tasos Mallios in numerous private communications.} In \cite{mall0}, for example, Tasos goes at great lengths, by using Gel'fand Duality and the so-called {\em Gel'fand Transform} of an algebra, in delimiting the sort of (topological) algebra sheaves that could be used as `good' structures sheaves $\struc$ in ADG. As Gel'fand Duality mandates, these are algebra sheaves over the suitably topologised spectrum of the original algebra.

At the same time, at the back of Tasos's mind, back in the day of the mid-90s when he was feverishly searching for solid founding pillars to erect ADG, must have surely been his beloved {\em \underline{Differential} Geometry} \cite{mall0}: that is to say, consciously or unconsciously (and here this author only speculates in retrospect), Tasos must have asked himself:

\noindent $\bullet$ {\em How can one extract \underline{differential geometric} structure (not just topological) from Algebra (:algebraic structure), thus in a sense emulate Gel'fand duality, but in a differential geometric setting?}

To that end, motivating inspiration must have come to Tasos from another celebrated mathematical result, which also elegantly depicts the categorical duality between Algebra and Geometry: {\em the Serre-Swan Theorem} \cite{serre, swan}, to which we briefly turn next.

\item \underline{\bf The Serre-Swan Theorem:} Jean-Pierre Serre's version of the theorem, which to this author is more pertinent to Tasos's original {\em differential geometric} endeavours and quests, roughly posits that {\em for every commmutative unital (Noetherian) ring $R$, then the category of finitely generated projective $R$-modules (Algebra) is equivalent to the category of algebraic vector bundles $\mathcal{V}$ ({\it i.e.}, locally free sheaves of structure sheaf $R$-modules of constant finite rank $n$) on the Spectrum $Spec(R)$ of $R$}.

The alert and astute reader, who is also familiar with the basic rudiments of ADG, must surely speculate that for Tasos, the result above must have come as an `epiphany moment' in his quest for algebraic structures to model ADG, if one substitutes:

\noindent $\bullet$ `{\em finitely generated differential $A$-modules}'\footnote{Where $A$ here is not just a ring, but is an algebra $A$ over a field ($\com$).} for `{\em finitely generated projective $R$-modules}' in the Serre-Swan Theorem; and also,

\noindent $\bullet$ `{\em vector sheaves $\modl$ ({\it i.e.}, locally free sheaves of structure sheaf $\struc$-differential modules of constant finite rank $n$) }' instead of `{\em  algebraic vector bundles $\mathcal{V}$ ({\it i.e.}, locally free sheaves of structure sheaf $R$-modules of constant finite rank $n$)}'.

\end{enumerate}

\noindent {\it In toto}, the usual $\smooth$-smooth vector bundles $\mathcal{V}$ on which the whole edifice of Classical Differential Geometry (CDG)---the so-called Classical Differential Calculus on Smooth (Differential) Manifolds---and its manifold applications to Modern Physics rest\footnote{We have empasised throughout our joint work with Tasos \cite{malrap1, malrap2, malrap3, malrap4} that CDG is a special case of, and can be recovered from, ADG when one assumes $\struc\equiv\smooth(M)$---that is, when one simply assumes copies of the algebra $\smooth(M)$ of $\smooth$-smooth functions on a differential (spacetime) manifold $M$ as occupying the stalks of the structure sheaf $\struc$ in the theory. Equivalently, it has been recently shown  that {\em the category $\mathcal{M}$ of $\smooth$-smooth manifolds is a full subcategory of the category $\mathcal{DT}$ of ADG-theoretic differential triads} \cite{demiran}.} is abstracted and generalised in ADG by {\em vector sheaves} $\modl$ ({\it i.e.}, locally free $\struc$-modules of finite rank $n$) over an in principle arbitrary topological space $X$ \cite{mall1, mall2, mall4}.\footnote{Recall from Section 1 that in our finitary case, the base space on which our finitary sheaves of differential incidence Rota algebras $\Omega_{i}$ are soldered is the very (primitive) spectrum $Spec(\Omega_{i})$ of those algebras---a `discrete' instance of {\em `Gel'fand Duality meets Serre-Swan'}.}

As an additional bonus at this point, the ADG-theoretic extension and generalisation of the Serre-Swan Theorem was the principal move of Tasos in \cite{mall5, mall6, mall4} towards arriving at a manifestly functorial geometric (pre)quantisation and second (:field) quantisation scheme for his ADG-theoretic field theory, with concomitant classification of the fields' elementary particle quanta into bosons and fermions, as (\ref{eq32}) above depicts.

\subsubsection{Tracing the Physical `Origins' of ADG: Breaking Algebra-Geometry Duality in Favour of an Algebraic Physical  Dynamics}

The two celebrated mathematical results mentioned above---Gel'fand Duality and the Serre-Swan Theorem---were seen to express a fundamental categorical duality (:functorial equivalence) between Algebra and Geometry and, arguably, they were speculated to be centers of inspiration and motivation for Professor Mallios in developing ADG as a mathematical abstraction, expansion, generalisation and enrichment\footnote{As Tasos originally preferred to call it cumulatively: {\em an axiomatisation} \cite{mall1, mall2}.} of the usual CDG on smooth manifolds.

In this subsection, we turn our attention to Tasos's fundamental {\em Physical} intuitions and motivations in applying ADG to what he used to call `{\em Physical Space(Time)}' and `{\em Physical Geometry}'. 

To that end, we quote directly from \cite{mall10} the following `definition' {\it \'a-la} Mallios of what he perceived as, and, {\it in extenso}, what ought to qualify as being, `{\em Physical Geometry}':

\begin{quotation}

\noindent ``{\em ...Physical Geometry is the `outcome' of the physical laws...}''

\end{quotation}

\noindent For which he then displayed the following `causal nexus' for {\em producing Physical Geometry from Dynamical Laws}:\footnote{Again, the quotation is borrowed {\it verbatim} from \cite{mall0}.}

\begin{quotation}

\noindent ``...Now, by looking at the technical correspondence/association,

\end{quotation}

$$\mathit{physical\, law}\longleftrightarrow\aconn-\mathit{connection,}$$

\begin{quotation}

\noindent one realizes that [the displayed expression above] might also be construed, as an
\emph{equivalent analogue of} the implication;

\end{quotation}

$$\aconn-\mathit{connection} (:\,\mathit{physical\, law})\Longrightarrow \mathit{curvature}\, (:\,\mathit{`geometry'},\, \mathrm{alias},\,
    \mathit{`shaping'})$$

\begin{quotation}

\noindent Consequently, still to repeat the above, but state it otherwise, one concludes that:

\end{quotation}

\centerline{ It is actually the \emph{physical laws}, that \emph{make}, what we might call (physical) \emph{`geometry'}....''}

\hskip 0.1in

\noindent All of Mallios's prophetic musings above may be subsumed under the following distilled {\em Fundamental Aphorism}:

\begin{quotation}

\noindent {\bf Fundamental Aphorism:} {\em Physical Space(Time)' and `Physical Geometry' is the result or the product of Field Dynamics}, in much the same way that, as we saw earlier in the paper, {\em the Curvature Field is a Geometric Morphism image of the Connection Field}.

\end{quotation}

\noindent In this line of thought, we conclude the present paper by quoting {\it verbatim} below the very opening paragraph of the wonderful {\em Exordium} in Mallios and Zafiris's last joint research monograph \cite{malzaf1}:\footnote{Some emphasised parts in the quote above are {\em our emphasis}.}

\begin{quotation}

\noindent ``{\em The major aim of the application of Analysis and Differential Geometry in Physics
is the setting up of a mechanism providing a precise description of the emergence
of geometric spectrums\footnote{That is to say, `{\em geometric spaces}'.} due to dynamical interactions, which can be further used
for making predictions}. In this manner, {\em the notion of a geometric spectrum is
considered as the outcome of a physical law or more generally of a dynamical interaction of a particular form}. 
This raises immediately the question if there exists
{\em an approach to physical geometric spectrums that is independent of any coordinate
point manifold background, in the sense that it refers directly to the physical relations causing the appearance of these spectrums without the intervention of any ad
hoc coordinate choices}. The answer provided to this question in this book is that
the theory of differential vector sheaves, that is geometric vector sheaves equipped
with a connectivity structure and obeying appropriate cohomological conditions,
provides the sought after functorial tool for a universal and natural approach to
physical geometric spectrums. The major difference of the proposed approach in
comparison to the traditional ones based on classical differential calculus and differential geometry of smooth manifolds consists in the realization that a classical
analytic technique is susceptible to a natural background-independent generalization if it is localizable by sheaf-theoretic means. In this case {\em the technique can be
expressed functorially, that is by means of natural transformations of sheaf functors
via the machinery of homological algebra}. {\em This is of crucial significance for setting
up a mechanism describing the emergence of physical geometric spectrums where
the notion of background smoothness is inapplicable}...''

\end{quotation}

\subsection{A Wish for the Future}

The whole time that I have known him, {\em Tasos was always with the underdog, ``always supporting and taking sides with the hunted, not the hunter''},\footnote{One of his own `proverbs of wisdom': ``{\em We are always with the hunted, not the hunter}''.} which he kept reminding me on a day-to-day basis. He also constantly urged me to {\em take risks} and be unconventional, unorthodox and iconoclastic in my research quests and endeavours in Quantum Gravity \cite{rap11}.

Thus, in paying tribute and homage to my wonderful teacher, mentor, friend and immortal companion in our joint Unending Quest, here's one of Tasos's favourite Feynman Quotes:\footnote{Taken from Richard Feynman's intro to his \cite{feyn}, where he talks about researchers taking the risk and `{\em going off into the wild blue yonder}' realm, seemingly strange and largely unexplored yet landscape, of Quantum Gravity.}

\begin{quotation}
\noindent ``...It is important that we don’t all follow the same fashion. We must increase the amount of
variety and the only way to do this is to implore you few guys, to take a risk with your own
lives so that you will never be heard of again, and go off into the wild blue yonder to see if you
can figure it out...”

\end{quotation}

\hskip 0.1in

{\em May his far reaching vision, the breadth of his conceptual perception, the imagination of his mathematical and physical intuition, the depth of his philosophical enquiry, the originality and the unorthodoxy of his approach, as well as the risk and the adventurousness of his research endeavours---all coupled to the priceless legacy that Professor Anastasios Mallios leaves behind---nurture, enrich, motivate and inspire future researchers in Quantum Gravity for years to come!}

\section*{Addendum I: An Anecdotal Exchange with Professor Mallios and a Conclusion Drawn from It}

I would like to share with this forum a private two-part exchange that we enjoyed with Tasos way back in May 1998, actually on the day of my 30th birthday, more than two years before his first 2-volume pitch of {\em Abstract Differential Geometry: The Geometry of Vector Sheaves} was published by Kluwer Academic Publishers \cite{mall1}. That exchange on the one hand dovetails perfectly with the two main mathematical results mentioned in the last section\footnote{{\em Gel'fand Duality and the Serre-Swan theorem}.} that, as I argued, must have played a pivotal role, consciously and/or unconsciously, in motivating and inspiring Tasos to develop ADG in the first place, and on the other, it casts light on Tasos's character as {\em a pure and young-at-heart, decent human being}.

I had just obtained my Ph.D. \cite{rap0}, with Professor Mallios as one of the external examiners of my thesis, and, very interested in the early developments of Topos Theory (TT) \cite{macmo}, I was naturally attracted by the core mathematical ideas and the working philosophy of one of TT's main early architects: {\em Alexandre Grothendieck} \cite{groth0}. I had also just finished reading the first part (:{\em Fatuity and Renewal}) of Grothendieck's `autobiographical' manuscript titled {\em Reaping and Sowing} \cite{groth} and we were discussing the wide range and far reaching depth of Alexandre Grothendieck's contributions to Modern Mathematics, especially to the field of Algebraic Geometry, via the introduction and application of novel Homological Algebra (:Category-theoretic) ideas, concepts and technical constructions.

{\bf Part 1 of the Exchange: Grothendieck's `Working Philosophy'} I remember I initiated the exchange by telling Tasos that I had just finished reading the first part of the {\em R\'ecoltes} and made the remark that the gist of Grothendieck's working philosophy in Mathematics was to attain an epoptic---as broad, as general and as abstract---viewpoint of the entire landscape of Mathematics. Tasos agreed and added two crucial ingredients:

\begin{enumerate}

\item That Grothendieck, willingly or not, explicitly or not, formally or informally (:intuitively), essentially {\em axiomatised} Mathematics; and,

\item That Grothendieck used to encounter and stare at the (only apparently) complex and esoteric Mathematics' landscape with {\em the innocence and the ignorance of a child}.

\end{enumerate}

\noindent $\bullet$ He backed the first point by saying that, much in the same vain as Grothendieck, he was planning to call (as he actually did!) his forthcoming work on `Abstract Differential Geometry: The Geometry of Vector Sheaves' {\em an Axiomatic Approach to Differential Geometry} \cite{mall1}. I asked him why did he think that the most epoptic, bird's eyeview of Mathematics could be attained by {\em Axiomatisation}, and he quoted me Aristotle from Nicomachean Ethics: ``{\it He who can properly define, divide and distinguish is to be considered God}'' \cite{arist}. He said that in Mathematics one can attain the broadest and most-encompassing viewpoint only when one has properly clarified and laid down the fundamental concepts and base axioms.

\noindent $\bullet$ He backed his second point by quoting back to me Bertrand Russell:\footnote{This is one of Tasos's Top-3 quotes.} ``{\em Men are born ignorant, not stupid. Education makes them stupid.}''. He said that Grothendieck had the gift and ability to {\em look at the World of Mathematics in innocent and ignorant awe and amazement}, with fresh eyes, not at all conditioned or biased by prior Knowledge or Education, but perhaps more importantly, {\em by not being afraid of making mistakes}. 

I indeed recall reading, a couple of days earlier, from the first part of Grothendieck's {\em R\'ecoltes} the following telling excerpt:

\begin{quotation}

\noindent ``{\itshape ...Discovery is the privilege of the child. It’s the little child that I want to talk about, the child who is not afraid to 
be wrong, to look silly, to not be serious, to not be like everyone else. He is neither afraid that the things he 
looks at will have a bad taste, different from what he expects, from what they appear to be, or rather: from what 
he has already understood them to be. He ignores the unspoken and unwavering consensus that form part of 
the air we breathe - which all the grown-ups are supposed to know and they do know. God knows (I suppose the 
grown-ups know well) if there have been any such child, since the dawn of ages!...

...The little child discovers the world as he breathes - the ebb and flow of his breath make him welcome the world 
in its delicate being, and makes him project himself into the world that also welcomes him. The adult can also 
discover, in those rare moments when he has forgotten his fears and his knowledge, when he looks at things or 
himself with eyes wide open, eager to know, new eyes - the eyes of a child...}''

\end{quotation}

{\bf Part 2 of the Exchange: Grothendieck's Work} With regard to Grothendieck's main contribution to Mathematics, especially in Algebraic Geometry, Tasos maintained that {\em Grothendieck essentially abstracted and purely algebraicised Algebraic Geometry, and effectively substituted Hard Analysis on `Rigid' Geometrical (:Arithmetic) Spaces by the more malleable and flexible inherently algebraic concepts and methods of Sheaf Cohomology}.

With the advent, blossoming and effective manifold applications of ADG to fundamental Mathematics and Physics, I can now draw confidently the following parallel in posthumous honour of Tasos:

\hskip 0.1in

\centerline{What {\bf Grothendieck} did for {\it Algebraic Geometry}, {\bf Mallios} did for {\it Differential Geometry}.}

\section*{Addendum II: Bohr's Poetic and Lexiplastic Imperative in Current and Future Quantum Gravity Research}

Preparing `psychologically' and `emotionally' the reader for the heuristic, yet technical and rigorous, Glossary that follows in the Appendix next, we recall and borrow almost {\it verbatim} from \cite{rap14},\footnote{From the very last section, titled {\em Poetry in Motion and in Action: the Future of Quantum Gravity Research}.} nearly two decades later(!), some still significant in our opinion remarks on {\em the importance of using `{\em (onomato)poetic language}', as well as novel conceptual (:theoretical/philosophical) and new technical  jargon, having manifest practical (:`calculational') implications and import, in our quest for a conceptually sound, philosophically cogent and technically creative and artful Quantum Theory of Gravity} (QG).

\paragraph{Descending to the quantum deep: the `experience-to-theoretical physics-to-mathematics-to-philosophy-to-poetry' ascension.}
In QG research, because of the glaring absence of experimental
data (in fact, of any prepared and controlled laboratory
experiments!)\footnote{Although at the same time, we are passive receptors of
cosmological data from the early universe.} to verify---or more
importantly, to falsify(!)---our theories, the
theoretical/mathematical physicist finds herself in the fortuitous
position of being free to roam in unconstrained, uninhibited
theory making, with sole guiding tools `aesthetic' elements such
as conceptual simplicity, economy, symmetry and beauty, backed by
mathematical abstraction, generality, rigor and logical
consistency. This has been appreciated as early as Dirac 
\cite{dirac3}, who, in trying to reason and evade singularities and unphysical infinities upon trying to quantise the electromagnetic field, 
implored theoretical physicists to explore and
use all the {\em mathematical} resources at their disposal, and
temporarily divesting experiments of their theory checking and
guiding role.

Ludwig Faddeev, for example, maintained fairly recently \cite{faddeev} that we
should finally break away from the classical theory-making 
route followed so far by theoretical physics, which can be
schematically represented by the cycle:

{\small
$$
\mathrm{experiments}~\mapto~\mathrm{predictions}\mapto\mathrm{mathematical~
formulations}\mapto\mathrm{further~experiments}
$$
}

\noindent and instead employ all our mathematical resources to
plough deeper into the foundations of `physical reality', leaving
experiments (and experimentalists!) to `catch up' with the new
mathematics (and with theoreticians!), not the other way around. In
this regard, we would like to borrow from \cite{faddeev} some
telling remarks made by Dirac from the aforementioned paper
\cite{dirac3}:\footnote{The quotation below is split into two
paragraphs (I and II), on which we comment separately after it.}

\begin{quotation}
\noindent {\bf Part I.} ``{\small ...The steady progress of physics requires for
its theoretical foundation a mathematics that gets continually
more advanced. This is only natural and to be expected. What,
however, was not expected by the scientific workers of the last
century was the particular form that the line of advancement of
the mathematics would take, namely, it was expected that the
mathematics would get more complicated, but would rest on a
permanent basis of axioms and definitions, {\em while actually the
modern physical developments have required a mathematics that
continually shifts its foundation and gets more abstract...It
seems likely that this process of increasing abstraction will
continue in the future and that advance in physics is to be
associated with a continual modification and generalization of the
axioms at the base of mathematics rather than with logical
development of any one mathematical scheme on a fixed
foundation.}\footnote{Our emphasis.} 

\noindent {\bf Part II.} There are at present fundamental problems in theoretical physics
awaiting solution [...]\footnote{Dirac here mentions a couple of
outstanding mathematical physics problems of his times. We have
omitted them.} the solution of which problems will presumably
require a more drastic revision of our fundamental concepts than
any that have gone before. Quite likely these changes will be so
great that it will be beyond the power of human intelligence to
get the necessary new ideas by direct attempt to formulate the
experimental data in mathematical terms. The theoretical worker in
the future will therefore have to proceed in a more indirect way.
{\em The most powerful method of advance that can be suggested at
present is to employ all the resources of pure mathematics in
attempts to perfect and generalise the mathematical formalism that
forms the existing basis of theoretical physics, and {\sl
after}\footnote{Dirac's own emphasis.} each success in this
direction, to try to interpret the new mathematical features in
terms of physical entities}\footnote{Again, our emphasis
throughout.}...}''
\end{quotation}

\begin{itemize}

\item {\bf Part I.} The words from this paragraph to be highlighted
with ADG-GT in mind here are: `{\em a mathematics that gets more
abstract}' and `{\em advance in physics is to be associated with a
continual process of abstraction {\rm [leading to a]} modification
and generalization of the axioms at the base of mathematics}'.
Indeed, the axiomatic ADG essentially involves an abstraction of
the fundamental notions of modern differential geometry ({\it e.g.},
connection), resulting in an entirely algebraic (:sheaf-theoretic)
modification and generalization of the latter's basic axioms
\cite{mall1, mall2, mall4}. And it is precisely this abstract and
generalized character of ADG that makes us hope that its
application could advance significantly (theoretical) physics, and
in particular, QG research. 

For, to quote again Einstein from earlier, in the
quantum deep we must look for ``{\em a purely algebraic method for
the description of reality}'' \cite{einst3}. \footnote{Alas, for
Einstein, the continuum spacetime and {\it in extenso} CDG-based
field theory was simply incompatible with the finitistic-algebraic
quantum theory \cite{stachel}, a divide that ADG has come a long
way to finally bridge
\cite{malrap1, malrap2, malrap3, malrap4, rap5, rap7, rap11, rap13, rap14}.}

\item {\bf Part II.} In this paragraph, apart from breaking from the
traditional cycle `experiment-theory-more experiment' mentioned
above ({\it i.e.}, Dirac's anticipation that `{\em new ideas {\rm
[won't come]} by direct attempts to formulate the experimental
data in mathematical terms}'), what should be highlighted is on
the one hand Dirac's prompting us `{\em to generalize the
mathematical formalism that forms the existing basis of
theoretical physics}', and on the other, `{\em to try to interpret
the new mathematical features in terms of physical entities}'.
Again, ADG goes a long way to fulfill Dirac's vision, since {\em the} (or at
least the bigger part of the) mathematics that lies at the heart
of current theoretical physics---namely, (the formalism of) {\em
differential geometry} ({\it i.e.}, the CDG on $\smooth$-smooth manifolds)---is
abstracted and generalized, while {\em after} this generalisation
has been achieved, the physical application and interpretation (of
ADG's novel concepts and features) has been carried out,
especially in the theoretical physics' field of quantum gauge
theories and gravity research. We believe that this is `{\em a
powerful method of advance}' indeed.

\end{itemize}

\noindent However, {\em this too is not enough} in our opinion. Existing mathematical
concepts, structures and techniques also come hand in hand with
implicit assumptions, hidden preconceptions and prejudices
associated with their historical development, {\it i.e.}, with past
problems other than QG(!) that they were invented in order to
formulate, tackle and (re)solve. Such preconceptions are very hard
to forget at the primary stages of theory making, let alone to
shed them altogether, especially when they have proved to be
experimentally successful in the past. Again Einstein, for
example, has given us a warning call regarding our almost
religious abiding by old, tried-and-tested concepts \cite{einst7}:

\begin{quotation}
\noindent ``{\small ...Concepts which have proved useful for
ordering things easily assume so great an authority over us, that
we forget their terrestrial origin and accept them as unalterable
facts. They then become labelled as `conceptual necessities', `a
priori situations', etc.\footnote{Think for instance of the
apparently fundamental notion of the `{\em spacetime continuum}':
``{\em time and space are modes by which we think, not conditions
in which we live}'' (as quoted by Manin in \cite{manin}). }The
road of scientific progress is frequently blocked for long periods
by such errors. It is therefore not just an idle game to exercise
our ability to analyze familiar concepts, and to demonstrate the
conditions on which their justification and usefulness depend, and
the way in which these developed, little by little...}"
\end{quotation}

For this, a few people have suggested to go even a bit further, past
mathematics, and into the realm of {\em Philosophy} to look for
novel QG research resources. 't Hooft, for example \cite{thooft},
insists that:

\begin{quotation}

\noindent ``{\small\em ...The problems of quantum gravity are much
more than purely technical ones. They touch upon very essential
philosophical issues...}''

\end{quotation}

\noindent For us, this will not suffice either. Philosophy too
comes burdened with a host of {\it a priori} concepts and
assumptions.\footnote{Especially the nowadays academic `Philosophy
of Science' \cite{sklar, clark}, which appears to be heavily
(almost paracytically!) dependent on the concepts, techniques,
results and current developments in science (and in particular, in
theoretical physics and applied mathematics).} 

Paraphrasing
Finkelstein in \cite{df}, ``{\em in the quantum deep one must travel
light}''. Alas, perhaps because of a deep psychological tendency
towards security (and an instinctive, biological one, for survival
\cite{wheat}), we tend to abide by what we already know and
(think) we understand (or believe to have a firm hold of backed by
numerous practical applications), and we take few `conservative
risks' (pun intended) towards standing bare, ignorant (but,
exactly thanks to this ignorance, uninhibited and unbiased!)
before Nature. This primordial fear of the unknown must be
overcome---at least it should be soothed by the Socratic stance
that, {\em anyway, the only thing that we know for sure is that we
know almost nothing}---and a way of achieving this is by engaging
into imaginative, creative {\em poetic activity} where there is
plenty of leeway for `trial-and-error' and a lot of room for
iconoclastic, unorthodox, unconventional and adventurous ideas that are
unburdened by ancestral theoretical demands or traditional beaten track 
conventions.

Indeed, granted that QG pushes us back to theorizing about the
archegonal acts of the World, what better means other than poetry
(with its analogies, metaphors and allegories) do we possess for
exploring, conceptually afresh and without {\it a priori}
commitments---ultimately, to deconstruct and reconstruct anew
\cite{plotnitsky}---the strange,\footnote{`Strange', of course,
relative to what we already (think we) know!} uncharted QG
landscape? 

Kandinsky's words echo ecophantically here \cite{kand}:

\vskip 0.1in

\centerline{``{\em Poetry brings us closer to the Creator.}''}

\vskip 0.1in

Especially regarding the unfamiliar realm of the quantum, we read
from \cite{midgley} (reading from \cite{tolstoy}):

\begin{quotation}
\noindent ``{\small ...In the first forty years of the twentieth
century, our vision of the physical world changed radically and
irretrievably. Atoms could behave like solid matter or like waves,
they were made of particles with strange top-like properties, with
nuclei which could disintegrate spontaneously, and, perhaps, set
up chains of disintegration themselves. For many, the most
interesting implication of all this new knowledge was, and still
is, philosophical. We have understood that our intuitive ideas of
what is possible and what is not---our common sense---are a result
of the conditioning of our minds by sense-experiences. {\em We
have had to change our ideas of what understanding consists
in}.\footnote{Midgley's emphasis.} As Bohr said, `{\sl When it
comes to atoms, language can only be used as in poetry. The poet,
too, is not nearly so concerned with describing facts as with
creating images}.'\footnote{Our {\sl emphasis}.} The same is true
of cosmological models, curved spaces and exploding universe. {\em
Images and analogies are the keys}.\footnote{Midgley's emphasis,
and mine.} Not you, not I, not Einstein could interpret the
universe in terms wholly related to our senses. Not that it is
incomprehensible, no. {\sl But we must learn to ignore our
preconceptions concerning space, time and matter, abandon the use
of everyday language and resort to metaphor. \underline{We must
try to think like poets}...''}\footnote{{\sl Emphasis} (and
underlining) is all ours.}}
\end{quotation}

\noindent What we have in mind here is that, in order to see and
tackle the problem of QG afresh, we must foremost be able to sort
of `(re)create it from scratch', forgetting for a while the
voluminous body of work---the various theoretical `evidence' that
different approaches to QG provide us with---that has been
gathered over the last 70+ years of research on it. The spirit of
Feynman comes to mind:

\vskip 0.1in \centerline{``{\em What I cannot create, I do not
understand.}'' \cite{feyn}\footnote{In the `{\em Quantum Gravity}'
prologue by Brian Hatfield.}} \vskip 0.1in

Of course, by `poetry' above all we mean {\em creation of new
conceptual terminology within a novel theoretical and technical
framework'}. 

In this respect, it is perhaps more important to stress that ADG
is not so much a {\em new} theory of DG---the main `{\em
mathematical formalism that forms the existing basis of
theoretical physics}', following Dirac's expression earlier---but
a theoretical framework that abstracts, generalises, revises and
recasts the existing CDG on differential manifolds by isolating and capitalising on its
fundamental, {\em essentially algebraic} (:`relational', in a
Leibnizian sense) features, which are not dependent at all on a
background locally Euclidean geometrical `space(time)'
(:manifold). In a way, from the novel viewpoint of ADG, we see
`old' and `stale' problems ({\it e.g.}, the $\smooth$-smooth singularities
of the manifold and CDG based GR) with `new' and `fresh' eyes \cite{rap5}.

Schopenhauer's words from \cite{schopenhauer} immediately spring
to mind:

\begin{quotation}
\noindent ``{\small\em ...Thus, the task is not so much to see
what no one has yet seen, but to think what nobody yet has thought
about that which everybody sees\footnote{All emphasis is
ours.}...}''
\end{quotation}

\paragraph{On the `idiosyncratic' terminology side.} As we will witness in the Glossary section following next, the novel perspective on gravity that ADG enables us to entertain
is inevitably accompanied by {\em new terminology}. We have thus
not refrained from engaging into vigorous poetic, `{\em
lexiplastic}' activity, so that our work in this paper abounds with
new, `idiosyncratic' terms for novel concepts hitherto not
encountered in the standard theoretical physics' jargon and
literature, such as `{\em gauge theory of the third kind}', `{\em
third quantisation}', `{\em synvariance}' and `{\em
autodynamics}', to name a few.

In this respect, we align ourselves with Wallace Stevens's words in
\cite{stevens}:

\vskip 0.1in

\centerline{``{\small\em ...Progress in any aspect is a movement
through changes in terminology...}''\footnote{Another one of Mallios's favourite quotes.}}

\vskip 0.1in

\noindent with the `changes in terminology' in our case being not
just superficial (:formal) `nominal' ones introduced  as it were
for `flash, effect and decor', but necessary ones coming from {\em
a significant change in basic theoretical framework for viewing
and actually doing DG in QG}: {\em from the usual geometrical manifold
based one (CDG), to the background manifoldless and purely
algebraic (:sheaf-theoretic) one of ADG}.

\paragraph{The bottom line is a verse: a Word for the World.} According to the Biblical Genesis, `{\em In the beginning was the Word}',
thus the ultimate task for future QG (re)search is to find the
right `words' to begin our theory making about the very beginning
of the World. For, to quote Bohr (as quoted in \cite{au}):

 \begin{quotation}
\noindent ``...{\small It is wrong to think that the task of
physics is to point out how nature is. {\em Physics concerns what
we can say about nature}}...''\footnote{Our emphasis. What could
baffle the reader here is the following apparent oxymoron: while
on the one hand we seem to advocate the aforesaid principle of
ADG-field realism (maintaining that the connection field $\conn$
exists `out there' independently of us experimenters,
measurers/geometers and theoreticians), on the other we endorse
Bohr's dictum above. Again, there's no paradox here: what {\em we}
can say about Nature ({\it ie}, in this case, about the field
$\conn$) is all encoded in the generalised arithmetics $\struc$
that {\em we} choose to represent it (on $\modl$). However, the
$\struc$-functoriality of the dynamics secures the independence of
the (dynamics of the) field from our generalized measurements (and
hence from our geometrical representations, {\it eg}, `spacetime')
in $\struc$ (and {\it in extenso} $\modl$, which is locally a
power of $\struc$).}
\end{quotation}

\noindent As Finkelstein notes,\footnote{In an early draft of
\cite{df} given to this author back in 1993.}

\begin{quotation}
\noindent ``{\small ...The fully quantum theory lies somewhere
within the theorizing activity of the human race itself, or the
subspecies of physicists, regarded as a quantum system. If this is
indeed a quantum entity, then the goal of knowing it completely is
a Cartesian fantasy, and {\em at a certain stage in our
development we will cease to be law-seekers and become
law-makers}.\footnote{For more discussion on this theme, see the
section in \cite{rap11}, titled `{\em The Saviors of Physical
Law}', emulating Kazantzakis' ``{\em The Saviors of God}''
\cite{kaz}. Our emphasis.}

It is not clear what happens to the concept of a correct theory
when we abandon the notion that it is a faithful picture of
nature. Presumably, just as the theory is an aspect of our
collective life, its truth is an aspect of the quality of our
life...}''
\end{quotation}

\noindent And what better means other than our Logos---or better,
than our imaginative and creative Logos: our poetic and bardic
{\em Mythos}---do we possess for approximating the archegonal
Truth about Nature? Moreover, what a humbling thought this is:
that in the end we may find out that this truth is the
quintessential quality of our ellogous lives. Then, in a
Nietzscheic sense \cite{nehamas}, {\em we will have become what we
already are: Poets true to our Nature!}

\section*{Appendix: Glossary of New Terminology and Heuristic ADG-GT Jargon}

In this concluding Appendix to the paper, we outline a Glossary of the novel ADG-theoretic terminology and conceptual heuristics that abound throughout this paper, plus of some that have made recurring appearances throughout our publications in the last two and a half decades \cite{malrap1,malrap2,malrap3,malrap4,rap1,rap2,rap5,rap7,rap11,rap13,rap14}.

The Glossary below is listed lexicographically, not in order of importance or frequency of appearance in past papers, and all the items are pre-fixed by `{\em ADG-}'.

\begin{enumerate}

\item {\bf ADG-$\struc$-Invariance.} The functorial imperative of ADG that the dynamical laws of physics (:here, Einstein's differential equations and Yang-Mills differential equations) should be respected by (:be `invariant' under) any of our (choices of) generalised coordinates or measurements (:arithmetics) employed in the structure sheaf $\struc$. As we saw in this paper, $\struc$-invariance entails local gauge invariance (see below).

\item {\bf ADG-Autodynamics.} The idea that the ADG-fields $\mathcal{F}=(\modl ,\conn)$, whether Maxwell, Yang-Mills or Einstein, are dynamically autonomous, `self-governing', `unitary', `holistic', `self-contained' entities, with no need for externally imposed spacetime parameters or `degrees of freedom' for their dynamical sustainance. 

\item {\bf ADG-$\cons$-Algebraized Space.} An in principle arbitrary topological space $X$ endowed with a structure sheaf $\struc$ of generalised arithmetics or `coordinates' localised on it: $(X,\struc)$.

\item {\bf ADG-Connection and Curvature Field Categories.} As we have the {\em Maxwell} $\mathcal{T}_{Max}=\{ (\mathcal{L},\conn_{Max})\}$, the {\em Yang-Mills} $\mathcal{T}_{YM}=\{ (\modl ,\conn_{YM})\}$, and the {\em Einstein} category $\mathcal{T}_{Einst}=\{ (\modl ,\conn_{Einst})\}$ of ADG-fields $(\modl ,\conn)$, we also define three corresponding {\em ADG-curvature field functor categories}: $\mathcal{C}_{Max}$, $\mathcal{C}_{YM}$ and $\mathcal{C}_{Einst}$, whose objects are ADG-curvature fields as in (\ref{eq24}), and whose arrows are \emph{natural transformation} type of correspondences between their $\otimes_{\struc}$-functorial objects.

\item {\bf ADG-Connection/Curvature Geometric Morphism.} The pair of adjoint functors $\mathcal{G}\mathcal{M}_{\struc}:=(\otimes_{\struc},\mathrm{Hom}_{\struc})$ effectuating functorial, \emph{natural transformation} type of correspondences between the category of ADG-fields and the corresponding category of ADG-curvature fields.

\item {\bf ADG-Curvature Space.} This is defined as the following quintet: $(\struc ,d , \Omega^{1}, d ,\Omega^{2})\equiv (\struc ,\conn, \Omega^{1},\conn^{2} ,\Omega^{2})$, consisting of a differential triad and a $d/\conn$-extension of the sheaf $\Omega^{1}$ of differential $1$-forms to a sheaf $\Omega^{2}$ of differential $2$-forms so as to be able to define the curvature of a connection according to (\ref{eq17}).

\item {\bf ADG-Curvature Field.} A pair consisting of a structure sheaf $\struc_{X}$ on a $\cons$-algebraized space $X$ and the curvature $\curv(\conn)$ of an $\struc$-connection $\conn$ (on a vector sheaf $\modl$) acting as an $\struc$-morphism: $\mathcal{R}=(\struc , \curv(\conn))$.

\item {\bf ADG-Differential Triad.} A triplet consisting of a structure sheaf $\struc_{X}$ on some $\cons$-algebraized space $X$, and a flat $\cons$-linear Leibnizian connection $d$ acting as a $\cons$-linear sheaf morphism that maps $\struc_{X}$ to a sheaf $\Omega$ of differential $\struc$-modules on $X$: $\triad=(\struc_{X} ,d ,\Omega(X))$

\item {\bf ADG-Field.} A pair consisting of a vector sheaf $\modl$ and an $\struc$-connection $\conn$ on it acting as a $\cons$-linear Leibnizian sheaf morphism: $\mathcal{F}=(\modl ,\conn)$.

\item {\bf ADG-Field Quantal Self-Duality.} The basic intuitive-heuristic observation that for any ADG-field $\mathcal{F}=(\modl ,\conn)$, (the local sections of) $\modl$ represent(s) some abstract kind of local quantum particle `position' states, while the action of the connection field $\conn$ on them represents some kind of generalised `momentum' type of action. The two structures are said to be `quantum complementary' aspects of the `unitary' and `coherent' ADG-field in the sense that they obey some abstract (sheaf cohomological) commutation Heisenberg uncertainty relations which define 3rd Quantisation in our scheme.

\item {\bf ADG-Field Solipsism/Monadology.} The idea that the ADG-fields $\mathcal{F}=(\modl ,\conn)$ are the sole dynamical entities (:variables) in our theory---the sole {\em physical} entities in our ADG-GT---without any `spacetime realm and reality' external to and separate from them. In this sense, the ADG-Field Solipsism is tantamount to {\em the ADG-Field Pure Realism},\footnote{We borrow the Tractarean idea of Ludwig Wittgenstein from \cite{witt}, that: ``{\em Solipsism coincides with Pure Realism}''.} namely, that the auto-dynamical, self-governing and self-transforming physical laws that the ADG-fields define and obey in-themselves are invariant no matter what, independently of what, structure group sheaf $\struc$ of generalised arithmetics---however reticular, pathological or singular---we employ to localise, coordinatise or `measure' them. In this sense,{\em  the ADG-fields are `physically real' entities} \cite{malrap3, malrap4, rap7, rap13, rap14}. This is another manifestation of Mallios's Principle of $\struc$-Invariance.

\item {\bf ADG-Gauge Theory of the 3rd Kind.} The idea that the ADG-field dynamics remains invariant under the `gauge' group of dynamical self-transmutations $\aut_{\struc}\modl$ of the ADG-fields $\mathcal{F}=(\modl ,\conn)$. It follows that {\em all symmetries and invariances of the ADG-field dynamics are internal to} ({\it i.e.}, happen within) {\em the fields themselves}, without recourse or reference to an external (:background) spacetime manifold. There is no external (:spacetime) versus internal (:gauge) symmetries' distinction in our theory. {\em All transformations are pure gauge transformations, in the sense that they are changes in the generalised coordinate gauges (:arithmetics) in} $\struc$, without reference to an external spacetime.

\item {\bf ADG-Gel'fand Duality.} The general idea that `differentiable space' comes from the structure sheaf $\struc$ of our generalised arithmetics.

\item {\bf ADG-Geometric Space.} The general idea that `the geometry of physical space' comes from `algebraic (:relational) dynamics' obeyed by, the dynamical relations between, the ADG-fields.

\item {\bf ADG-Natural Transformation.} This pertains to the {\em Natural Transformation} character of the fundamental {\em Geometric Morphism} $\mathcal{G}\mathcal{M}_{\struc}:=(\otimes_{\struc},\mathrm{Hom}_{\struc})$ between the corresponding functor categories of ADG-Connection Fields and ADG-Curvature Fields (or the ADG-Curvature Spaces that the latter define). This is another expression of Mallios's Principle of $\struc$-Invariance.

\item {\bf ADG-Principle of $\struc$-Algebraic Relativity of Differentiability.} Since {\em all differentiability in ADG derives from the structure sheaf $\struc$ of algebras of arithmetics or `generalised coordinates'}, different choices of $\struc$ entail different `differential geometric mechanisms' (:`Calculus'), but the dynamical laws of Nature---the very differential equations that can be formulated via that differential geometric mechanism that these $\struc$s define---remain invariant under them. This is yet another expression of Mallios's Principle of $\struc$-Invariance.

\item {\bf ADG-Synvariance.} The ADG-theoretic analogue of (General) Covariance in accord with ADG-Autodynamics above; namely, that in much the same way that $\mathrm{Diff}(M)$---the group of active diffeomorphisms of the `external' base spacetime manifold of GR---represents the Principle of General Covariance (PGC) of GR, $\aut_{\struc}\modl$---the group of $\struc$-automorphisms of the vector sheaf $\modl$---represents the invariance group of dynamical self-transmutations of the Einstein ADG-field $\mathcal{F}_{Einst}=(\modl ,\conn_{Einst})$. {\it Mutatits mutandis} then for $\mathcal{F}_{Max}$ and $\mathcal{F}_{YM}$.

\item {\bf ADG-`Unitary' Quantal Gauge Field.} This pertains to a tetrad of functorially and dynamically closely entwined structures $\mathbf{U}:=(\modl ,\conn, \aut_{\struc}\modl , \mathcal{Q})$, and it subsumes under a single coherent and inseparable `unitary whole' all the four most important functorial structural traits of ADG-GT, namely: {\em `local quantum particle states' represented by local sections of a vector sheaf $\modl$, their `dual-complementary' functorial ADG-gauge field dynamics generated by an algebraic $\struc$-connection $\conn$, the latter's local gauge invariance of the 3rd kind encoded in the principal structure sheaf $\aut_{\struc}\modl$ of $\modl$'s automorphisms, and the dual particle-field canonical-type of 3rd quantisation, represented by the functorial morphism $\mathcal{Q}$ between the relevant sheaf categories involved}.

\end{enumerate}

\section*{Acknowledgments}

After a hiatus of one-and-a-half decades, this paper comes as a result of {\em serendipitous privilege} and {\em dogged perseverance}: serendipitous privilege for crossing worldlines and working closely with {\em Tasos Mallios}, and dogged perseverance from having read, on the year that Tasos departed, {\em Charles Bukowski}'s poem {\it Go All the Way}. Both gentlemen taught me in their own inimitable way that ``{\em Isolation is the Gift}'' \cite{bukowski}, for which I am eternally indebted to them. 

\hskip 0.1in

\noindent The unceasing `moral' support of my lovely family: {\em Kathleen, Francis, James} and {\em Cookie}, is also lovingly aknowledged, especially their patience and understanding in putting up with me over the years.

\end{document}